\documentclass{jpc}

\usepackage{multicol}
\usepackage{array}
\usepackage{bigstrut}
\usepackage{listings}
\usepackage{color}
\usepackage[dvipsnames,usenames]{xcolor}
\usepackage{booktabs}
\usepackage{url}
\usepackage{xspace}
\usepackage{amssymb}
\usepackage{amsmath}
\usepackage{amsthm}
\usepackage{listings}
\usepackage{color}
\usepackage{courier}
\usepackage[font=small,labelfont={bf}]{subcaption}
\usepackage[font=small,labelfont={bf}]{caption}
\usepackage{graphicx}
\usepackage{ifpdf}
\usepackage[T1]{fontenc}
\usepackage{ae,aecompl}
\usepackage{rotating}
\usepackage{sepnum}
\usepackage[above,below]{placeins}
\usepackage{balance}
\usepackage{csquotes}
\usepackage{wrapfig}
\usepackage{framed}

\usepackage{mathtools}

\definecolor{darkred}{rgb}{0.75,0,0}

\newcommand{\point}[1]{\par\smallskip\noindent{\small\bf #1:}}

\newcommand\TODO[1]{\textcolor{red}{TODO: {#1}}}
\newcommand{\tool}{\textsc{Blender}\xspace}

\newcommand{\E}{\mathrm{E}}
\newcommand{\Var}{\mathrm{Var}}
\newcommand{\Cov}{\mathrm{Cov}}

\makeatletter
\def\url@foostyle{%
  \@ifundefined{selectfont}{\def\UrlFont{\sf}}{\def\UrlFont{\small\ttfamily}}}
\makeatother

\newcommand{\empirical}[1]{{#1}}
\newcommand{\tr}[1]{{}}

\setlist[itemize,1]{leftmargin=5mm}

\usepackage{algorithm}
\usepackage[noend]{algpseudocode}
\algdef{SE}[SUBALG]{Indent}{EndIndent}{}{\algorithmicend\ }%
\algtext*{Indent}
\algtext*{EndIndent}
\makeatletter
\algrenewcommand\ALG@beginalgorithmic{\footnotesize}
\makeatother
\newcommand{\params}{\vspace{2mm}\textsf{Parameters}\par\smallskip\footnotesize}
\newcommand{\variables}{\normalsize\textsf{Variables}\par\smallskip\footnotesize}

\newcommand{\etal}{\emph{et.~al.}\xspace}

\makeatletter
\newtheorem*{rep@theorem}{\rep@title}
\newcommand{\newreptheorem}[2]{%
\newenvironment{rep#1}[1]{%
 \def\rep@title{#2 \ref{##1}}%
 \begin{rep@theorem}}%
 {\end{rep@theorem}}}
\makeatother
\newreptheorem{theorem}{Theorem}
\newreptheorem{observation}{Observation}

\let\svthefootnote\thefootnote
\newcommand\blankfootnote[1]{%
	\let\thefootnote\relax\footnotetext{#1}%
	\let\thefootnote\svthefootnote%
}

\definecolor{javared}{rgb}{0.6,0,0} 
\definecolor{javagreen}{rgb}{0.25,0.5,0.35} 
\definecolor{javapurple}{rgb}{0.5,0,0.35} 
\definecolor{javadocblue}{rgb}{0.25,0.35,0.75} 

\lstdefinelanguage{JavaScript}{
	keywords={POSE,GESTURE,EXECUTION,and,SUCCESS,SCORE,APP},
	keywordstyle=\color{javablue}\bfseries,
	ndkeywords={left,right,degrees,counter,public,clockwise,frontal,on,the,plane,hand,arm,wrist,elbow,rotate,return,with,your,point,down,up,seconds,slowly,fast,stretch,hold,new,for,seconds,touch,static},
	keywordstyle=\color{javared}\bfseries,
	ndkeywordstyle=\color{blue}\bfseries,
	numberstyle=\color{green}\bfseries,
	otherkeywords={:, \,, ., $, \{, \}, \[, \]},
	basicstyle=\scriptsize\ttfamily,
	identifierstyle=\color{black},
	sensitive=false,
	comment=[l]{//},
	morecomment=[s]{/*}{*/},
	commentstyle=\color{darkgray}\ttfamily,
	stringstyle=\color{javared}\ttfamily,
	numberstyle=\color{javagreen}\ttfamily,
	morestring=[b]',
	morestring=[b]",
	alsoletter=-,
}

\lstdefinestyle{numbers} {
	numbers=left,
	stepnumber=1,
	numbersep=8pt,
	numbersep=10pt,
	xleftmargin=.5\parindent,
	frame=single,
	basicstyle=\ttfamily\footnotesize
}

\makeatletter
\AtBeginDocument{%
	\let\c@figure\c@lstlisting
	
	\let\ftype@lstlisting\ftype@figure 
}
\makeatother

\begin{document}

\title[BLENDER: Enabling Local Search with a Hybrid Differential Privacy Model]{BLENDER: Enabling Local Search with\\a Hybrid Differential Privacy Model}
\titlecomment{This paper is published in the \textit{Journal of Privacy and Confidentiality, Vol.~9 (2) 2019}. The preliminary version appeared at the \textit{26th USENIX Security Symposium} in 2017~\cite{blender}.\\ \textsuperscript{*}Supported in part by NSF grant \#1755992 and a gift from Mozilla.}

\author[B.~Avent]{Brendan Avent\textsuperscript{*}}
\address{University of Southern California}
\email{bavent@usc.edu}

\author[A.~Korolova]{Aleksandra Korolova\textsuperscript{*}}
\address{University of Southern California}
\email{korolova@usc.edu}

\author[D.~Zeber]{David Zeber}
\address{Mozilla}
\email{dzeber@mozilla.com}

\author[T.~Hovden]{Torgeir Hovden}
\address{Mozilla}
\email{torgeir@eritreum.com}

\author[B.~Livshits]{Benjamin Livshits}
\address{Imperial College London}
\email{livshits@ic.ac.uk}

\def\abstractname{Abstract}
\newcommand{\nnum}[1]{\sepnum{.}{,}{}{#1}}
\newcommand{\ttime}[1]{\sepnum{.}{,}{}{#1}}
\newcommand{\hide}[1]{}
\newcommand{\callbackExample}[2]{{#1 (#2):}}
\newcommand{\code}[1]{{\ifmmode{\mathtt{#1}}\else$\mathtt{#1}$\fi}}
\newcommand{\htmltag}[1]{\code{<\!\!#1\!\!>}}
\newcommand{\smallcode}[1]{{\footnotesize\ifmmode{\mathtt{#1}}\else$\mathtt{#1}$\fi}}
\newcommand{\verysmallcode}[1]{{\scriptsize\ifmmode{\mathtt{#1}}\else$\mathtt{#1}$\fi}}
\newcommand{\verysmallcodebf}[1]{{\scriptsize\ifmmode{\bm{\mathtt{#1}}}\else$\bm{\mathtt{#1}}$\fi}}

\makeatletter
\renewcommand{\verbatim@font}{\footnotesize\fontfamily{cmtt}\selectfont}

\newtheorem{theorem}{Theorem}[section]
\newtheorem{definition}[theorem]{Definition}
\newtheorem{observation}[theorem]{Observation}

\newcounter{example}
\renewcommand{\theexample}{\arabic{example}}
\newenvironment{ex}{
\refstepcounter{example}{
\vskip -7pt \noindent {\bf \\ Example \theexample \ }}}{ \ $\Box$
}

\newenvironment{verbacode}[2]{%
    \newcommand*{\mycaptiontext}{#1}%
    \newcommand*{\mylabeltext}{#2}%
    \begin{figure*}[tb]%
    \begin{centering}%
    \rule[2pt]{\linewidth}{0.6pt}
    \begin{minipage}{12cm}%
    \scriptsize%
    \begin{alltt}}{%
    \end{alltt}%
    \end{minipage}%
    \par%
    \end{centering}%
    \vspace*{2pt}%
    \rule[10pt]{\linewidth}{0.6pt}
        \vspace*{-6ex}%
    \caption{\mycaptiontext}%
    \label{\mylabeltext}%
    \end{figure*}}

\newenvironment{columnfig}[3]{%
    \newcommand*{\mycaptiontext}{#2}%
    \newcommand*{\mylabeltext}{#3}%
    \begin{figure}[#1]%
    \begin{centering}%
    \rule[2pt]{\linewidth}{0.6pt}
    \begin{minipage}{\columnwidth}}{%
    \end{minipage}%
    \par%
    \end{centering}%
    \vspace*{2pt}%
    \rule[10pt]{\linewidth}{0.6pt}
        \vspace*{-6ex}%
    \caption{\mycaptiontext}%
    \label{\mylabeltext}%
    \end{figure}}


\makeatletter

\newcommand{\zug}[1]{\langle #1 \rangle}
\newcommand \Pre   {\ensuremath{\mathit{Pre}}}
\newcommand \Post  {\ensuremath{\mathit{Post}}}
\newcommand \Prog  {\ensuremath{\mathsf{Prog}}}
\newcommand \Spec  {\ensuremath{\mathsf{Spec}}}
\newcommand \Impl  {\ensuremath{\mathsf{Impl}}}
\newcommand \Path  {\ensuremath{\mathsf{Path}}}
\newcommand \Triple{\ensuremath{\mathsf{Triple}}}
\newcommand \PreE  {\ensuremath{\mathit{PreE}}}
\newcommand \PostE {\ensuremath{\mathit{PostE}}}
\newcommand \wpc   {\ensuremath{\mathsf{WP}}}
\newcommand \eimpl {\ensuremath{\Rrightarrow}}
\newcommand \expify[1] {\lbrack #1 \rbrack}

\newcommand \xsrc  {\ensuremath{X_{\mathit{src}}}}
\newcommand \xsnk  {\ensuremath{X_{\mathit{src}}}}
\newcommand \xsan  {\ensuremath{X_{\mathit{src}}}}

\newcommand{\ab}[1]{\textbf{(AB:#1)}}

\begin{abstract}
We propose a \emph{hybrid} model of differential privacy that considers a combination of regular and opt-in users who desire the differential privacy guarantees of the local privacy model and the trusted curator model, respectively. We demonstrate that within this model, it is possible to design a new type of \textit{blended} algorithm that improves the utility of obtained data, while providing users with their desired privacy guarantees.
We apply this algorithm to the task of privately computing the head of the search log and show that the blended approach provides significant improvements in the utility of the data compared to related work.
Specifically, on two large search click datasets, comprising~\empirical{1.75} and~\empirical{16}~GB, respectively, our approach attains NDCG values exceeding~\empirical{95\%} across a range of privacy budget values.
\end{abstract}

\maketitle

\section{Introduction}
\label{sec:intro}
Now more than ever organizations are confronted with the tension between collection and sharing of mass-scale user data to fuel innovations and user's privacy. Today, an organization that needs user data to improve the quality of its service often has no choice but to perform its own data collection.
However, its users may not want to share their raw data with the organization, especially if they consider it to be sensitive. Furthermore, by collecting this user data, the organization assumes liability, as it may be leaked through security breaches, required to be shared through subpoenas, or indirectly leaked by the output of computations done on the data. Thus, both organizations and users would benefit not only from strong, rigorous privacy guarantees of the data sharing and use, but also from minimizing the raw data collected by the organization to achieve their goal. Thus, data collection with differential privacy in the local model is the best match for user expectations of privacy and the guarantees an organization may want to provide. However, due to the amounts of data required in order to achieve meaningful utility when ensuring privacy in the local model, such collection is relevant only to the biggest organizations with massive user bases, and the smaller ones get edged out. The goal of this work is to open possibilities for use of differential privacy to include organizations with smaller user bases. \\

\point{Local differential privacy}
Over the last several years, we have seen some examples of the local differential privacy~(LDP) model beginning to be used for data collection in practice, most notably in the context of the Chrome web browser~\cite{erlingsson2014rappor} and Apple's data collection~\cite{wired}.

In the LDP model, the data collector (such as Google or Apple) obtains aggregate data statistics without observing the exact values of user's private data. This is achieved by applying a privacy-preserving perturbation to each user's raw data before it leaves the user's device. 
This approach protects not only the individual users, but also the data collector from risks such as data breaches. \\

\point{Trusted curator model}
An alternative model that has been most commonly used in the academic literature on differential privacy to date is the \emph{trusted curator model}, where a curator first collects each user's private data and then produces and releases a privacy-preserving version of the collected dataset.
In this model, although users are guaranteed that the released dataset protects their privacy, they must be willing to share their private, unperturbed data with the curator and trust that the curator properly performs a privacy-preserving perturbation. \\

\point{Hybrid model}
The contribution of this paper stems from our observation that the two models can co-exist. 
People's attitudes toward privacy vary widely~\cite{acquisti2005privacy,acquisti2015privacy,dienlin2015privacy}, and some users may be comfortable with sharing their data with a trusted curator, while others may require the privacy protections of the local model. 

In industry practice, many companies already rely on a group of beta testers with whom they have higher levels of mutual trust.
It is not uncommon for such beta testers to voluntarily opt-in to a less privacy-preserving model than that of an average end-user~\cite{microsoft-opt-in}. 
For example, Mozilla warns potential beta users of its Firefox browser that ``Pre-release versions automatically send Telemetry data to Mozilla to help us improve Firefox\footnote{\url{https://www.mozilla.org/en-US/privacy/firefox/}}''; Microsoft states that ``[Windows Insider Program] services may automatically collect and provide data to Microsoft, which may include your personal information\footnote{\url{https://insider.windows.com/en-us/program-agreement/}}''; Google has a similar provision for the beta testers of the Canary build of the Chrome browser\footnote{\url{https://www.chromium.org/getting-involved/dev-channel}}.

For these users (referred to as the \emph{opt-in group}), the trusted curator privacy model is a natural match. 
For all other users (referred to as \emph{clients}), the local privacy model is appropriate.
Our goal is to demonstrate that by separating the user pool into these two groups, according to their trust (or lack thereof) in the data aggregator, we can improve the utility of the collected data while preserving privacy.
We dub this new model the \textit{hybrid differential privacy} model. \\

\point{Applications}
We consider two specific applications in this paper to demonstrate the usefulness of the hybrid model: \emph{local search} provided by a browser and \emph{search trend computation}.

Local search revolves around the problem of how a browser maker can collect information about users' clicks as they interact with search engines\footnote{A browser maker may choose to combine the results obtained from user interactions that stem from several search engines depending on the context or surface results obtained from Baidu and not Bing depending on the user's current location.} in order to create the \emph{head} of the search logs, i.e., the collection of the most popular queries and their corresponding URLs, to be made available to users \textit{locally} (i.e., on their devices). Specifically, it involves computing on query-URL pairs, where the URLs are those clicked as a result of submitting the query and receiving a set of answers. With proper privacy measures in place, the head of the search logs can then be deployed in the end-user browser to serve the most common queries with a very low latency or in situations when the user is disconnected from the network.

Local search can also be thought of as a form of caching, where many queries are answered in a manner that does not require a round trip to the server. Such local caching of the most frequently posed search queries has a disproportionately positive impact on the expected query latency~\cite{silvestri2010mining,baeza2007impact}, as search engine queries follow a power-law distribution~\cite{baeza2008design}. \\

Search trend computation entails finding the most popular queries and sorting them in order of popularity. An example of this is the Google trends service\footnote{\url{https://www.google.com/trends/}}, which has an up-to-date list of trending topics and queries.\\

\point{Utility challenges}
Local search and search trend computation can be thought of as problems in the category of \emph{heavy hitter discovery and estimation}, which is a well-studied problem in the context of information retrieval. 
Heavy hitter discovery is also one of the canonical problems in privacy-preserving data analysis~\cite{bhaskar2010discovering, privbasis}.
Moreover, the recent work in the LDP model is focused on precisely that problem~\cite{erlingsson2014rappor, qin2016heavy, fanti2016building} or very closely related ones of histogram computations~\cite{bassily2015local, kairouz2016discrete}. However, current privacy-preserving approaches in the local model lead to utility losses that are quite significant, to a point where the results are no longer useful for local search. 
For instance, it is common to seek NDCG~\cite{jarvelin2002cumulated, valizadegan2009learning} values above 0.9 for the local search problem of finding the most popular queries; however, the current best algorithm applied to this problem under the LDP model~\cite{qin2016heavy} is only able to attain an NDCG value of 0.385 while ensuring LDP with an $\epsilon$ of $5$ (see Section~\ref{sec:ccs} for further detail).

If privacy constraints make the utility too low compared to the original, the privacy-preserving approach is at risk to not be adopted.
This is especially true in the context of search tasks, where users have been conditioned for years to expect high-quality results.

\subsection{Contributions}
Our work makes the following contributions:
\begin{itemize}
    \item Introduces and utilizes a realistic, hybrid trust model, which removes the traditional ``all-or-nothing'' trust assumption towards a central curator.
    \item Proposes \tool, an algorithm that takes advantage of the hybrid differential privacy model for computing heavy hitters. Specifically, \tool utilizes data obtained from the opt-in users in order to modify the privacy-preserving algorithm run for all other users and then combines the data of opt-in and all other users in an informed way, in order to improve the utility of the privacy-preserving computation.
    \item Performs a comprehensive utility evaluation of \tool on two large search click datasets, comprising~\empirical{1.75} and~\empirical{16}~GB for two applications: search trend computation and local search. Demonstrates that \tool achieves high levels of utility (i.e., NDCG values in excess of~\empirical{95\%}) while maintaining differential privacy for reasonable privacy parameter values.
    \item Provides the first empirical demonstration that hybrid trust models, such as those combining data provided in the local model of differential privacy with data provided in the trusted curator model, can lead to non-trivial improvements in utility. Thus, it suggests the exploration of algorithms for such models as a promising direction for increasing the feasibility of differential privacy's deployment by both a wider range of organizations as well as for a wider variety of applications.
\end{itemize}

\section{Overview}
\label{sec:overview}
We now discuss the curator models that will form the basis of our hybrid model in more detail, provide a high-level overview of our proposed algorithm, \tool, that coordinates the privatization, collection and aggregation of data in this model, and discuss some of the specific choices we make in this algorithm.
We use the application of enabling local search based on user search histories while preserving differential privacy throughout; but, as will become clear from the discussion, our approach can be applied to other frequency-based discovery and estimation tasks.
 
\subsection{Differential Privacy and Curator Models} 
In the last decade, we have witnessed scores of ad-hoc approaches that have turned out to be inadequate for protecting privacy.
The problem stems from the impossibility of foreseeing all attacks of adversaries capable of utilizing outside knowledge.
Differential privacy, which has become the gold standard privacy guarantee in the academic literature, and is gaining traction in industry and government~\cite{erlingsson2014rappor, wired, machanavajjhala2008privacy}, overcomes the prior issues by focusing on the \textit{privatization} algorithm applied to the data, requiring that it preserves privacy in a mathematically rigorous sense under an assumption of an omnipotent adversary.

Most differentially private algorithms developed to date~\cite{dwork2014algorithmic} operate in the \textit{trusted curator model}: all users' private data is collected by the curator before privatization techniques are applied to it.
This means that although the privacy of the eventual result of the computation is ensured, the curator gets to observe the users' private data. 
However, as was most recently argued by Apple~\cite{wired}, users may not trust the data collector with their data, and may prefer privatization to occur before their data reaches the collector.
This is known as the \textit{local model}, since privatization occurs locally.

Although it may seem counter-intuitive, it is possible to obtain useful insights even when the data collector does not have access to the original data and receives only data that has already been locally privatized.
Suppose a data collector wants to determine the proportion of the population that is HIV-positive.
The local privatization algorithm works as follows:
each person contributing data secretly flips a biased coin.
If the coin lands heads, they report their true HIV status; otherwise, they report a status at random.
This algorithm, known as \textit{randomized response}~\cite{warner1965randomized}, guarantees each person plausible deniability and is differentially private (with privacy parameters determined by the bias of the coin).
But since the randomness is incorporated into the algorithm in a precisely specified way, the data collector is able to recover an estimate of the true proportion of HIV-positive people if enough people contribute their locally privatized data.\\

\begin{displayquote}
\textbf{Current differential privacy literature considers the trusted curator model and the local model entirely independently.
Our goal is to show that there is much to be gained by combining the two.\\}
\end{displayquote}

Formally, an algorithm $\mathcal{A}$ is $(\epsilon, \delta)$-differentially private~\cite{dwork2006calibrating} if and only if for all neighboring databases $D$ and $D^\prime$ differing in precisely one user's data, the following inequality is satisfied for all possible sets of outputs $Y \subseteq Range(\mathcal{A})$:
\vspace{1mm}
$$\Pr[\mathcal{A}(D) \in Y] \le e^\epsilon \Pr[\mathcal{A}(D^\prime) \in Y] + \delta.$$
\vspace{-3mm}

The definition of what it means for an algorithm to preserve differential privacy is the same for both the trusted curator model and the local model.
The only distinction is in the timing of when the privacy perturbation needs to be applied -- in the local model, the data needs to undergo a privacy-preserving perturbation before it is sent to the aggregator, whereas in the trusted curator model the aggregator may first collect all the data, and then apply a privacy-preserving perturbation.
The timing distinction leads to differences in what is meant by ``neighboring databases" in the definition and to differences in which algorithms are analyzed.
In the local model, $D$ represents data of a single user and $D'$ represents data of the same user, with possibly changed values.
In the trusted curator model, $D$ represents data of all users and $D^\prime$ represents data of all users, except one of the user's values may be altered.
Concretely, for the case of collecting a single search record from each user, the databases in the trusted curator model contain a collection of search records and differ in the value of one record, while the databases in the local model contain one record each.

\subsection{An Algorithm for the Hybrid Model}
As discussed in Section~\ref{sec:intro}, we consider two groups of users: the opt-in group, who are comfortable with privacy as ensured by the trusted curator model, and the clients, who desire the privacy guarantees of the local model.
Our proposed algorithm, \tool, coordinates the privatization, collection, and aggregation of the data from the opt-in and the client users.

\subsubsection{Outline of Our Approach}
The core of our innovation is to take advantage of the privatized information obtained from the opt-in group in order to create a more efficient (in terms of utility) algorithm for data collection from the clients.
Furthermore, the privatized results obtained from the opt-in group and from the clients are then ``blended" in a way that takes into account the privatization algorithms used for each group, and thus, again, achieving an improved utility over a less-informed combination of data from the two groups.

The problem of enabling local search using past search histories can be viewed as the task of identifying the most frequent search records among the population of users, and estimating their underlying probabilities (both in a differential privacy-preserving manner).
In this context, we call the data collected from the users \textit{search records}, where each search record is a pair of strings of the form $\langle query, URL\rangle$, representing a query that a user posed followed by the URL that the user subsequently clicked.
We denote by 
$p_{\langle q, u\rangle}$ the true underlying probability of the search record $\langle q, u\rangle$ in the population.
We assume that our algorithm receives a sample of users from the population, each holding their own collection of private data drawn independently and identically from the distribution over all records $p$.
Its goal is to output an estimate $\hat{p}$ of probabilities of the most frequent search records, while preserving differential privacy (in the trusted curator model) for the opt-in users and (in the local model) for the clients.\\

\begin{wrapfigure}{r}{6.0cm}
\vspace{-4mm}
\centering
\ifpdf
\includegraphics[height=0.8\textheight]{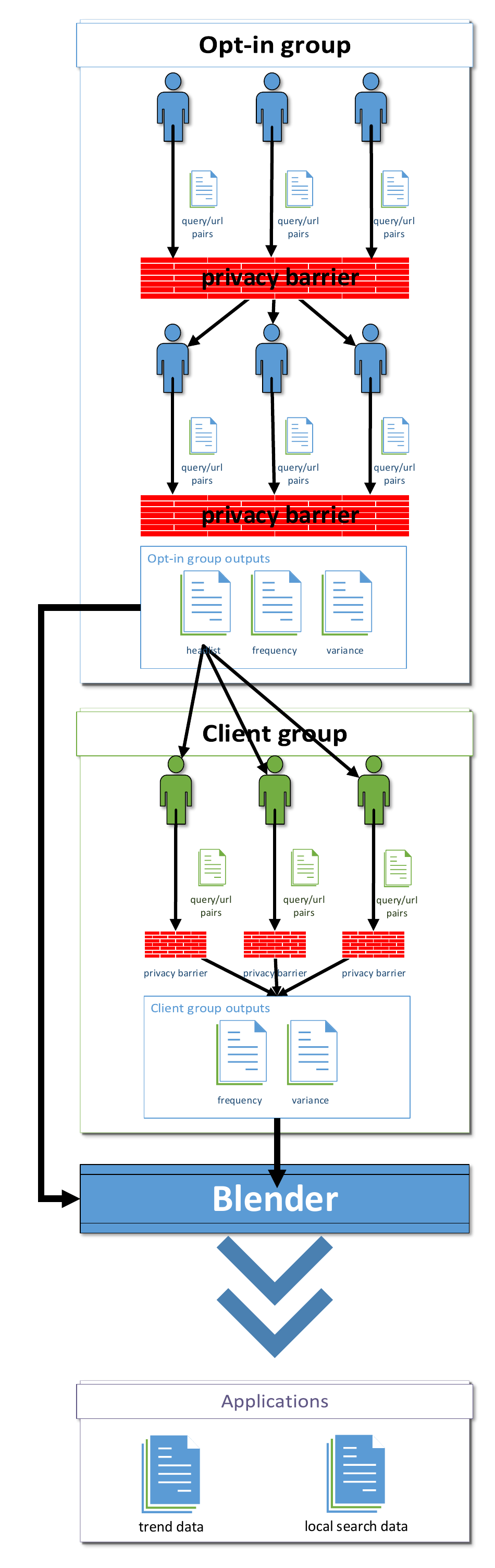}
\fi
\caption{Architectural diagram of \tool's processing steps.}
\label{fig:arch}
\vspace{-4mm}
\end{wrapfigure}

\point{Informal Overview of \textsc{Blender}}
Figure~\ref{fig:arch} presents an architectural diagram of \tool. 

The core of our approach is in utilizing the strengths of each of the models.
Specifically, the \emph{head list discovery} portion of the task -- that is, finding the names of the most-frequent queries and URLs -- can be done much more effectively under the trusted curator model than under the local model.
Thus, we assign most opt-in users to this task.
With the domain significantly narrowed, the remaining users are then assigned to the \emph{frequency estimation} portion of the task, where the underlying frequencies of the queries and URLs are estimated.

\tool serves as the trusted curator for the opt-in group of users, and begins by aggregating data from them.
Using a portion of the data, it constructs a candidate ``head list'' of records in a differentially private manner that approximates the most common search records in the population.
It additionally includes a single ``wildcard'' record, $\langle \star,\star \rangle$, which represents all records in the population that weren't previously included in the candidate head list.
It then uses the remainder of the opt-in data to estimate the probability of each record in the candidate head list in a differentially private manner, then (optionally) trims the candidate head list down further creating the final head list.
This result of this component of the algorithm is the privatized trimmed head list of search records and their corresponding probability and variance estimates, which can be shared with each user in the client group, as well as with the world.

Each member of the client group receives the head list obtained from the opt-in group.
Each client then individually uses the head list to apply a differential privacy-preserving perturbation to their data, subsequently reporting their perturbed results to \tool.
\tool then aggregates all the clients' reports and, using a statistical denoising procedure, estimates both the probability for each record in the head list as well as the variance of each of the estimated probabilities based on the clients' data.

Finally, for each record, \tool combines the record's probability estimates obtained from the two groups.
It does so by taking a convex combination of the groups' probability estimates using their respective variance estimates.
\tool outputs the obtained records and their combined record estimates, which can then be used to drive local search, determine trends, and more.\\

\begin{figure}[h!]
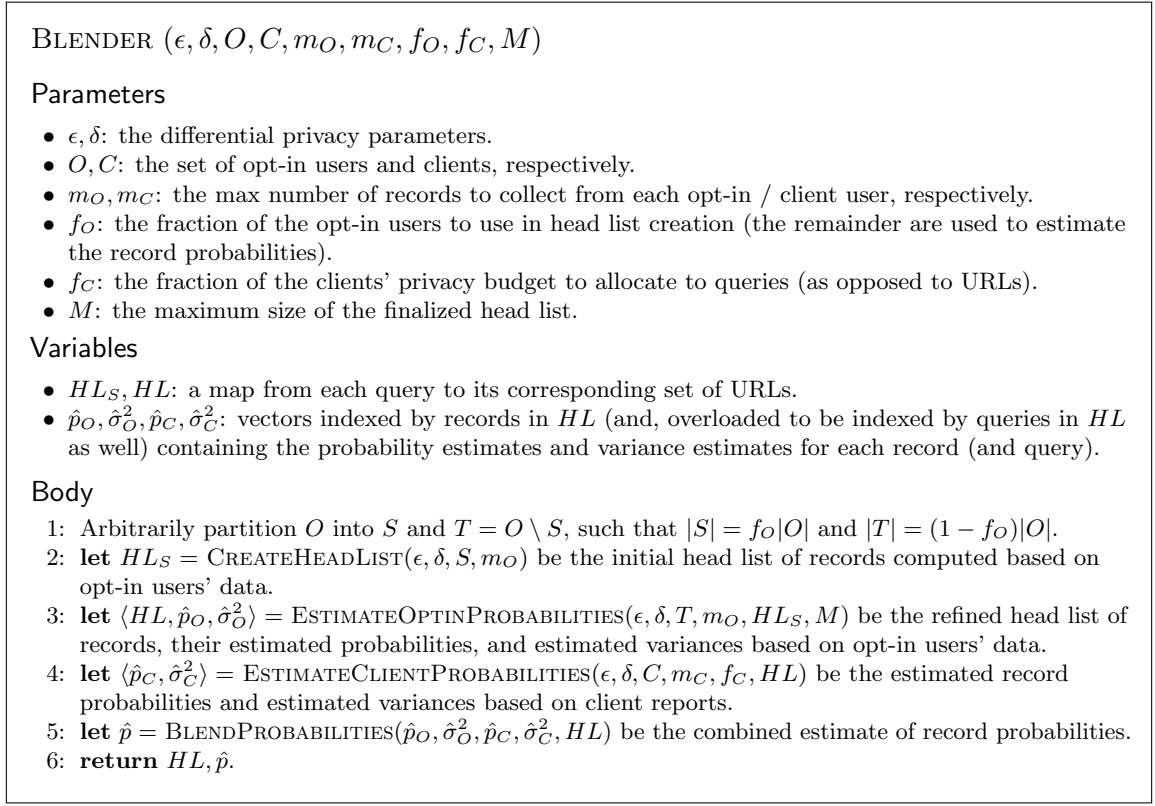

\raggedright
\begin{framed}
\begin{algorithmic}[1]{
\params
\begin{itemize}
	\item $\epsilon, \delta$: the differential privacy parameters. 
	\item $O, C$: the set of opt-in users and clients, respectively.
	\item $m_O, m_C$: the max number of records to collect from each opt-in  /  client user, respectively.
	\item $f_O$: the fraction of the opt-in users to use in head list creation (the remainder are used to estimate the record probabilities).
	\item $f_C$: the fraction of the clients' privacy budget to allocate to queries (as opposed to URLs).
	\item $M$: the maximum size of the finalized head list.
\end{itemize}
\variables
\begin{itemize}
	\item $HL_S, HL$: a map from each query to its corresponding set of URLs.
	\item $\hat{p}_O, \hat{\sigma}^2_O, \hat{p}_C, \hat{\sigma}^2_C$: vectors indexed by records in $HL$ (and, overloaded to be indexed by queries in $HL$ as well) containing the probability estimates and variance estimates for each record (and query).
\end{itemize}
}
{\tool($\epsilon, \delta, O, C, m_O, m_C, f_O, f_C, M$)}
	\State Arbitrarily partition $O$ into $S$ and $T = O \setminus S$, such that $|S|=f_O |O|$ and $|T| = (1-f_O)|O|$.
	\Let $HL_S = $ \textproc{CreateHeadList}$(\epsilon, \delta, S, m_O$) be the initial head list of records computed based on opt-in users' data.
	\Let $\langle HL, \hat{p}_O, \hat{\sigma}^2_O \rangle =$ \textproc{EstimateOptinProbabilities}$(\epsilon, \delta, T, m_O, HL_S, M$) be the refined head list of records, their estimated probabilities, and estimated variances based on opt-in users' data.
	\Let $\langle \hat{p}_C, \hat{\sigma}^2_C \rangle =$ \textproc{EstimateClientProbabilities}{$(\epsilon, \delta, C, m_C, f_C, HL)$} be the estimated record probabilities and estimated variances based on client  reports.
	\Let $\hat{p} = $ \textproc{BlendProbabilities}{$(\hat{p}_O, \hat{\sigma}^2_O, \hat{p}_C, \hat{\sigma}^2_C, HL)$} \label{state-combine} be the combined estimate of record probabilities.
	\State \Return $HL, \hat{p}$.
\end{algorithmic}
\end{framed}
\caption{\tool, the server algorithm that coordinates the privatization, collection, and aggregation of data from all users.}
\label{fig:server-alg}
\end{figure}

\point{A Formal Overview of \textsc{BLENDER}\xspace}
Figure~\ref{fig:server-alg} presents the precise algorithmic overview of each step, including key parameters.
Lines 1-3 of \tool describe the treatment of data from opt-in users, line 4 -- the treatment of clients, and line 5 -- the process for combining the probability estimates obtained from the two groups.
The only distinction between opt-in users and clients in terms of privacy guarantees provided is the curator model -- trusted curator and local model, respectively.
Other than that, both types of users are assumed to desire the same level of $(\epsilon, \delta)$-differential privacy.

We will detail our choices for the privatization sub-algorithms and discuss their privacy properties next.
A key feature of \tool, however, is that its privacy properties do not depend on the specific choices of the sub-algorithms.
That is, the post-processing property of differential privacy \cite{dwork2014algorithmic} guarantees that as long as \textproc{CreateHeadList, EstimateOptinProbabilities}, and \textproc{EstimateClientProbabilities} each satisfy $(\epsilon, \delta)$-differential privacy in its respective curator model, then so does \tool.
This allows changing the sub-algorithms if better versions (utility-wise or implementation-wise) are discovered in the future.
Among the parameters of \tool, the first four (the privacy parameters and the sets of opt-in and client users) can be viewed as given externally, whereas the following five (the number of records collected from each user and the distribution of the privacy budget among the sub-algorithms' sub-components) can be viewed as knobs the designer of \tool is at liberty to tweak in order to improve the overall utility of \tool's results.

\subsubsection{Overview of \tool Sub-Algorithms}
We now present the specific choices we made for the sub-algorithms in \tool. Detailed technical discussions of their properties follow in Section~\ref{sec:technical}.\\

\begin{figure}[tb]
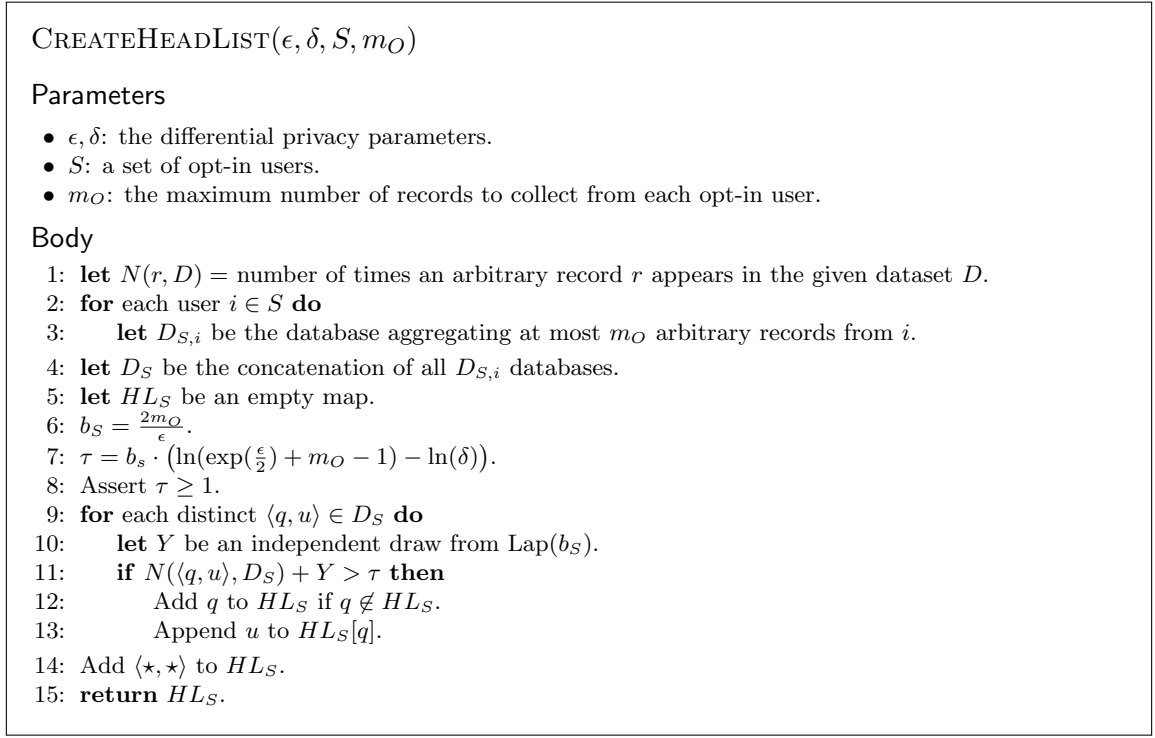

\raggedright
	\begin{framed}
		\begin{algorithmic}[1]{
				\params
				\begin{itemize}
					\item $\epsilon, \delta$: the differential privacy parameters.
					\item $S$: a set of opt-in users.
					\item $m_O$: the maximum number of records to collect from each opt-in user.
				\end{itemize}
			}
			{\textsc{CreateHeadList$(\epsilon, \delta, S, m_O)$}}
			\Let $N(r, D) =$ number of times an arbitrary record $r$ appears in the given dataset $D$.
			
			\For{each user $i \in S$}
				\Let $D_{S,i}$ be the database aggregating at most $m_O$ arbitrary records from $i$.
			\EndFor
			\Let $D_S$ be the concatenation of all $D_{S,i}$ databases.
			\Let $HL_S$ be an empty map.
			\State $b_S = \frac{2 m_O}{\epsilon}$.
			\State $\tau = \max \{ b_s \cdot \left(\ln(\exp(\frac{\epsilon}{2})+m_O-1)-\ln(\delta)\right)$, 1\}.
			\For{each distinct $\langle q,u\rangle \in D_S$}
				\Let $Y$ be an independent draw from \textrm{Lap}$(b_S)$.
				\If {$N(\langle q,u\rangle, D_S) + Y > \tau$}
					\State Add $q$ to $HL_S$ if $q \not \in HL_S$.
					\State Append $u$ to $HL_S[q]$.
				\EndIf
			\EndFor
			\State Add $\langle \star,\star\rangle$ to $HL_S$.
			\State \Return $HL_S$.
		\end{algorithmic}
	\end{framed}
	\caption{Algorithm for creating the head list from a portion opt-in users in a privacy-preserving way.}
	\label{fig:opt-in-alg-headlist}
	\vspace{-0mm}
\end{figure}

\begin{figure}[tb]
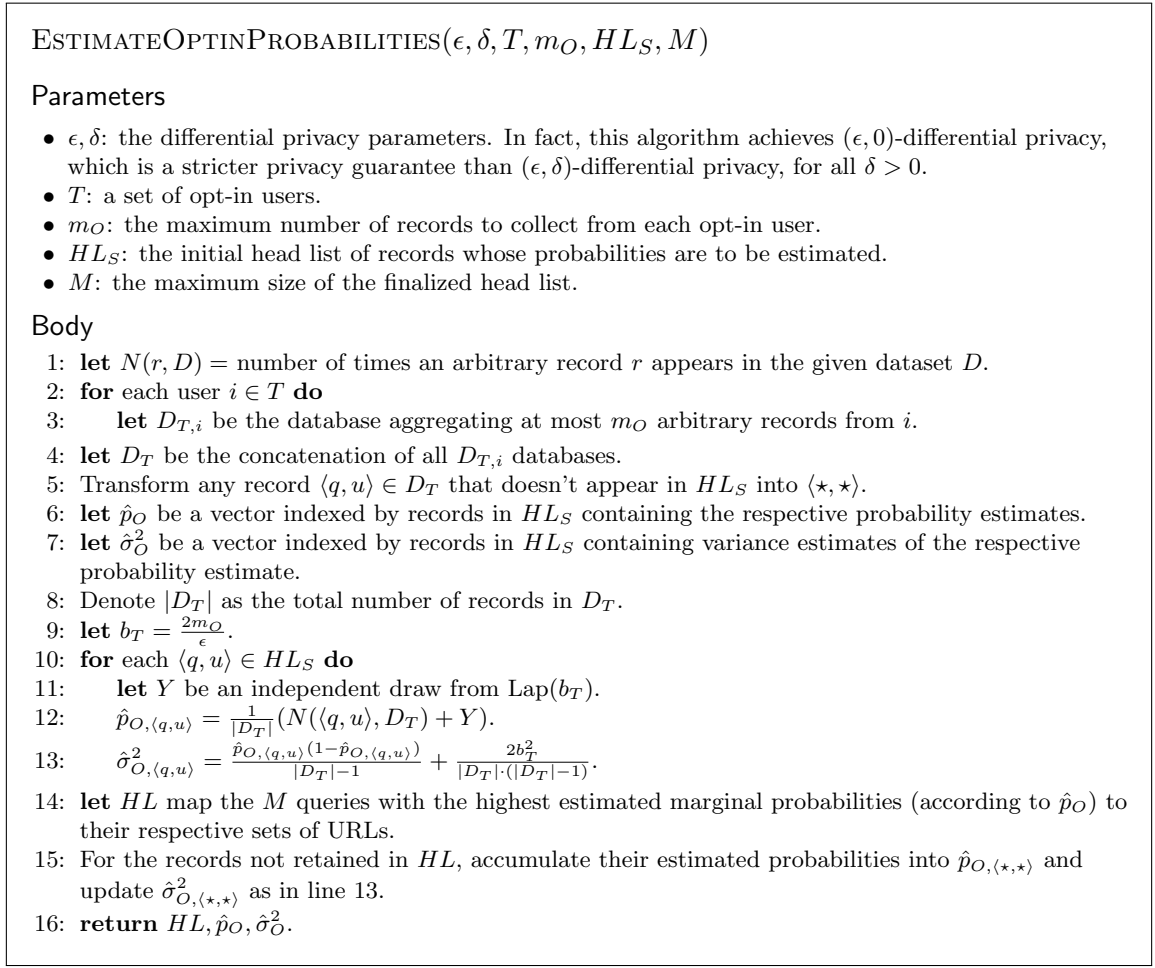

\raggedright
	\begin{framed}
		\begin{algorithmic}[1]{
				\params
				\begin{itemize}
					\item $\epsilon, \delta$: the differential privacy parameters. In fact, this algorithm achieves $(\epsilon, 0)$-differential privacy, which is a stricter privacy guarantee than $(\epsilon, \delta)$-differential privacy, for all $\delta>0$.
					\item $T$: a set of opt-in users.
					\item $m_O$: the maximum number of records to collect from each opt-in user.
					\item $HL_S$: the initial head list of records whose probabilities are to be estimated.
					\item $M$: the maximum size of the finalized head list.
				\end{itemize}
			}
			{\textsc{EstimateOptinProbabilities$(\epsilon, \delta, T, m_O, HL_S, M)$}}
			\Let $N(r, D) =$ number of times an arbitrary record $r$ appears in the given dataset $D$.
			
			\For{each user $i \in T$}
			\Let $D_{T,i}$ be the database aggregating at most $m_O$ arbitrary records from $i$.
			\EndFor
			\Let $D_T$ be the concatenation of all $D_{T,i}$ databases.
			\State Transform any record $\langle q,u\rangle \in D_T$ that doesn't appear in $HL_S$ into $\langle \star,\star\rangle$.
			\Let $\hat{p}_O$ be a vector indexed by records in $HL_S$ containing the respective probability estimates.
			\Let $\hat{\sigma}^2_O$ be a vector indexed by records in $HL_S$ containing variance estimates of the respective probability estimate.
			\State Denote $|D_T|$ as the total number of records in $D_T$.
			\Let $b_T = \frac{2m_O}{\epsilon}$.
			\For{each $\langle q,u\rangle \in HL_S$}
			\Let $Y$ be an independent draw from \textrm{Lap}$(b_T)$.
			\State  $\hat{p}_{O,\langle q,u\rangle} = \frac{1}{|D_T|}(N(\langle q,u\rangle, D_T) + Y)$.
			\State $\hat{\sigma}^2_{O,\langle q,u\rangle} = \frac{\hat{p}_{O,\langle q,u\rangle}(1-\hat{p}_{O,\langle q,u\rangle})}{|D_T|-1} + \frac{2b_T^2}{|D_T| \cdot (|D_T|-1)}$.
			\EndFor
			\Let $HL$ map the $M$ queries with the highest estimated marginal probabilities (according to $\hat{p}_O$) to their respective sets of URLs.
			\State For the records not retained in $HL$, accumulate their estimated probabilities into $\hat{p}_{O,\langle \star,\star\rangle}$ and update $\hat{\sigma}^2_{O,\langle \star,\star\rangle}$ as in line 13.
			\State \Return $HL, \hat{p}_O, \hat{\sigma}^2_O$.
		\end{algorithmic}
	\end{framed}
	\caption{Algorithm for privacy-preserving estimation of probabilities of records in the head list from a portion of opt-in users.}
	\label{fig:opt-in-alg-frequencies}
	\vspace{-0mm}
\end{figure}

\point{Algorithms for Head List Creation and Probability Estimation Based on Opt-in User Data (Figures~\ref{fig:opt-in-alg-headlist},~\ref{fig:opt-in-alg-frequencies})}
The opt-in users are partitioned into two sets -- $S$, whose data will be used for initial head list creation, and $T$, whose data will be used to estimate the probabilities and variances of records from the formed initial head list.

The initial head list creation algorithm, described in Figure~\ref{fig:opt-in-alg-headlist}, constructs the list in a differentially private manner using search record data from group $S$.
The algorithm follows the strategy introduced in~\cite{korolova2009releasing} by aggregating the records of the opt-in users from $S$, and including those records whose noisy count exceeds a threshold in the head list.
The noise to add to the true counts\footnote{\textrm{Lap}$(b)$ refers to a random draw from the Laplace distribution with scale $b$.} and the threshold are calibrated to ensure differential privacy, using~\cite{korolova2012protecting}.
The goal of the algorithm is to approximate the true set of most frequently searched and clicked search records as closely as possible, while ensuring differential privacy.

Our algorithm differs from previous work in two ways: 1) it replaces the collection and thresholding of queries with the collection and thresholding of records (i.e., query - URL pairs) and 2) its definition of neighboring databases is that of databases differing in one user's record values, rather than in the removal of one user's data.
These distinctions necessitate the choice of $m_O = 1$ as well as higher values for noise and threshold than in~\cite{korolova2012protecting}.

For those records included in the initial head list set, the algorithm described in Figure~\ref{fig:opt-in-alg-frequencies} uses the remaining opt-in users' data (from set $T$) to differentially privately estimate each record's probability, denoted $\hat{p}_O$.
The $M$ most frequent records in $\hat{p}_O$ are retained to form the final head list.
This algorithm is the standard Laplace mechanism from the differential privacy literature~\cite{dwork2006calibrating}, with scale of noise calibrated to our definition of neighboring datasets.
Our implementation ensures $(\epsilon, 0)$-differential privacy, which is a more stringent privacy guarantee than for any non-zero $\delta$.
We need to set $m_O=1$ for the privacy guarantees to hold, because we treat data at the search record rather than query level.

Finally, the head list is passed to the client group, and the head list and its probability and variance estimates are passed to the \textsc{BlendProbabilities} step of \tool.

The choice of how to split opt-in users into the sub-groups of $S$ and $T$ and the choice of $M$ are unrelated to privacy constraints, and can be chosen by \tool's developer to optimize utility goals, as will be discussed in Section~\ref{sec:parameter-choices}.\\

\begin{figure}[h!]
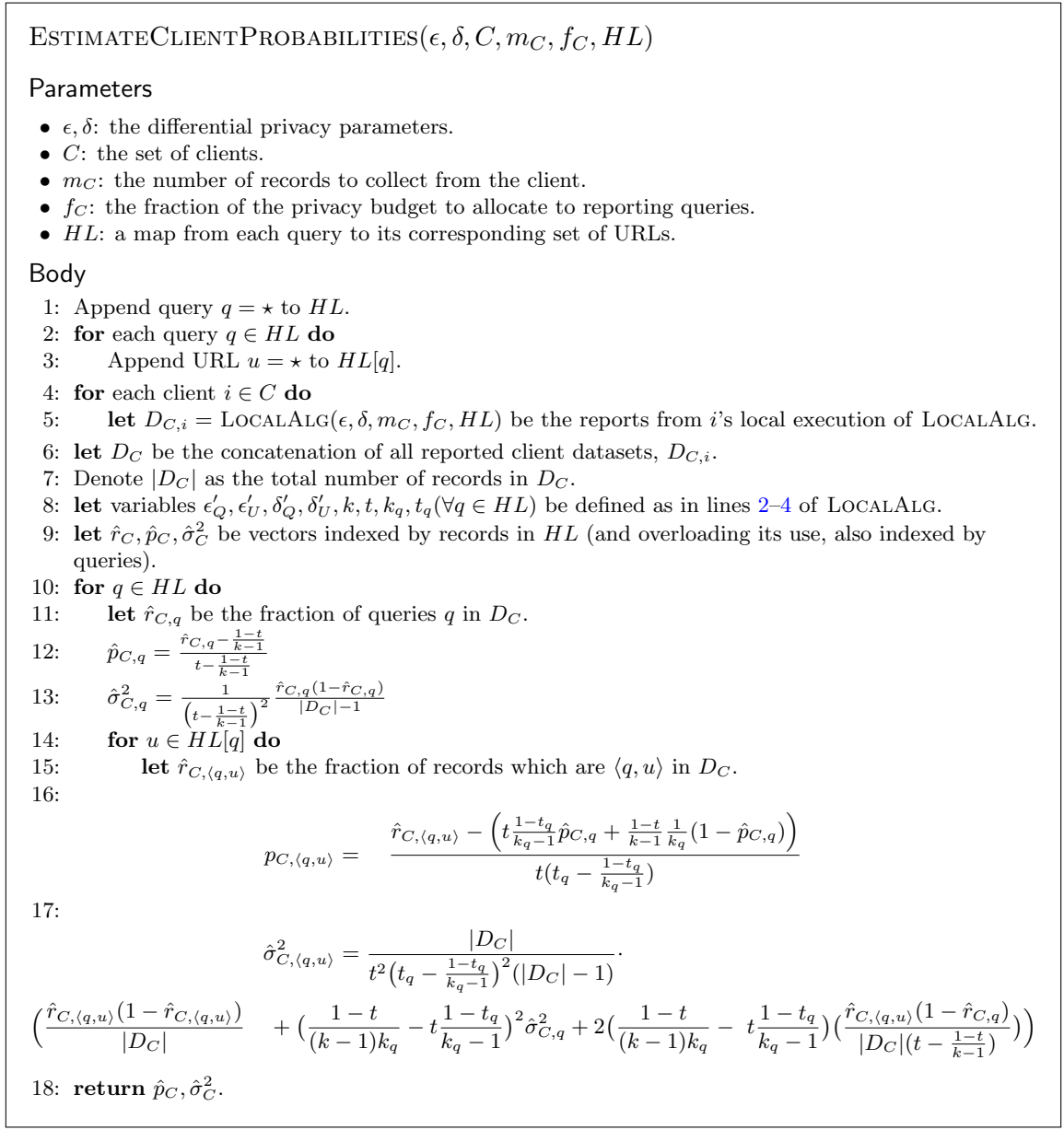

\raggedright 
\begin{framed}
\begin{algorithmic}[1]{
\params
\begin{itemize}
	\item $\epsilon, \delta$: the differential privacy parameters.
	\item $C$: the set of clients.
	\item $m_C$: the number of records to collect from the client.
	\item $f_C$: the fraction of the privacy budget to allocate to reporting queries.
	\item $HL$: a map from each query to its corresponding set of URLs.
\end{itemize}
}
{\textsc{EstimateClientProbabilities}$(\epsilon, \delta, C, m_C, f_C, HL)$}
	\State Append query $q=\star$ to $HL$.
	\For{each query $q \in HL$} 
		\State Append URL $u=\star$ to $HL[q]$.
	\EndFor
	
	\For{each client $i \in C$}
		\Let $D_{C,i} =$ \textproc{LocalAlg}$(\epsilon, \delta, m_C, f_C, HL)$ be the reports from $i$'s local execution of \textproc{LocalAlg}.
	\EndFor
	\Let $D_C$ be the concatenation of all reported client datasets, $D_{C,i}$.
	\State Denote $|D_C|$ as the total number of records in $D_C$.
	\Let variables $\epsilon^\prime_Q, \epsilon^\prime_U, \delta^\prime_Q, \delta^\prime_U, k, t, k_q, t_q (\forall q \in HL)$ be defined as in lines~\ref{begin-define-vars}--\ref{end-define-vars} of \textproc{LocalAlg}.
	\Let $\hat{r}_C, \hat{p}_C, \hat{\sigma}^2_C$ be vectors indexed by records in $HL$ (and overloading its use, also indexed by queries).
	\For{$q \in HL$}
		\Let $\hat{r}_{C,q}$ be the fraction of queries $q$ in $D_C.$
		\State $\hat{p}_{C,q} = \frac{\hat{r}_{C,q} - \frac{1-t}{k-1}}{t - \frac{1-t}{k-1}}$
		\State $\hat{\sigma}^2_{C,q} = \frac{1}{\bigl(t - \frac{1-t}{k-1}\bigr)^2} \frac{\hat{r}_{C,q}(1-\hat{r}_{C,q})}{|D_C|-1}$
		\For{$u \in HL[q]$}
			\Let $\hat{r}_{C,\langle q,u\rangle}$ be the fraction of records which are $\langle q,u\rangle$ in $D_C$.			
			\State \begin{align*}
			&{p}_{C,\langle q,u\rangle} = \quad \frac{\hat{r}_{C,\langle q,u\rangle} - \left(t\frac{1-t_q}{k_q-1}\hat{p}_{C,q} + \frac{1-t}{k-1}\frac{1}{k_q}(1-\hat{p}_{C,q})\right)}{t(t_q - \frac{1-t_q}{k_q-1})}
			\end{align*}
			
			\State \begin{align*}
			&\texttt{\phantom{x}\hspace{32mm}}\hat{\sigma}^2_{C,\langle q,u\rangle} = \frac{|D_C|}{t^2\bigl(t_q - \frac{1-t_q}{k_q-1}\bigr)^2 (|D_C|-1)} \cdot \\ 
			&\Bigl( \frac{\hat{r}_{C,\langle q,u\rangle}(1-\hat{r}_{C,\langle q,u\rangle})}{|D_C|}
			\quad + \bigl(\frac{1-t}{(k-1)k_q} - t\frac{1-t_q}{k_q-1} \bigr)^2 \hat{\sigma}^2_{C,q} + 2 
			\bigl(\frac{1-t}{(k-1)k_q} -~t\frac{1-t_q}{k_q-1} \bigr)\bigl(\frac{\hat{r}_{C,\langle q,u\rangle}(1-\hat{r}_{C, q})}{|D_C| (t-\frac{1-t}{k-1})} \bigr) \Bigr)
			\end{align*}
		\EndFor
	\EndFor
	\State \textbf{return} $\hat{p}_C, \hat{\sigma}^2_C$.
\end{algorithmic}
\end{framed}
\caption{Algorithm for estimating probabilities of records in the head list from the locally privatized reports of the client users.}
\label{fig:client-alg}
\vspace{-0mm}
\end{figure}

\begin{figure}[h!]
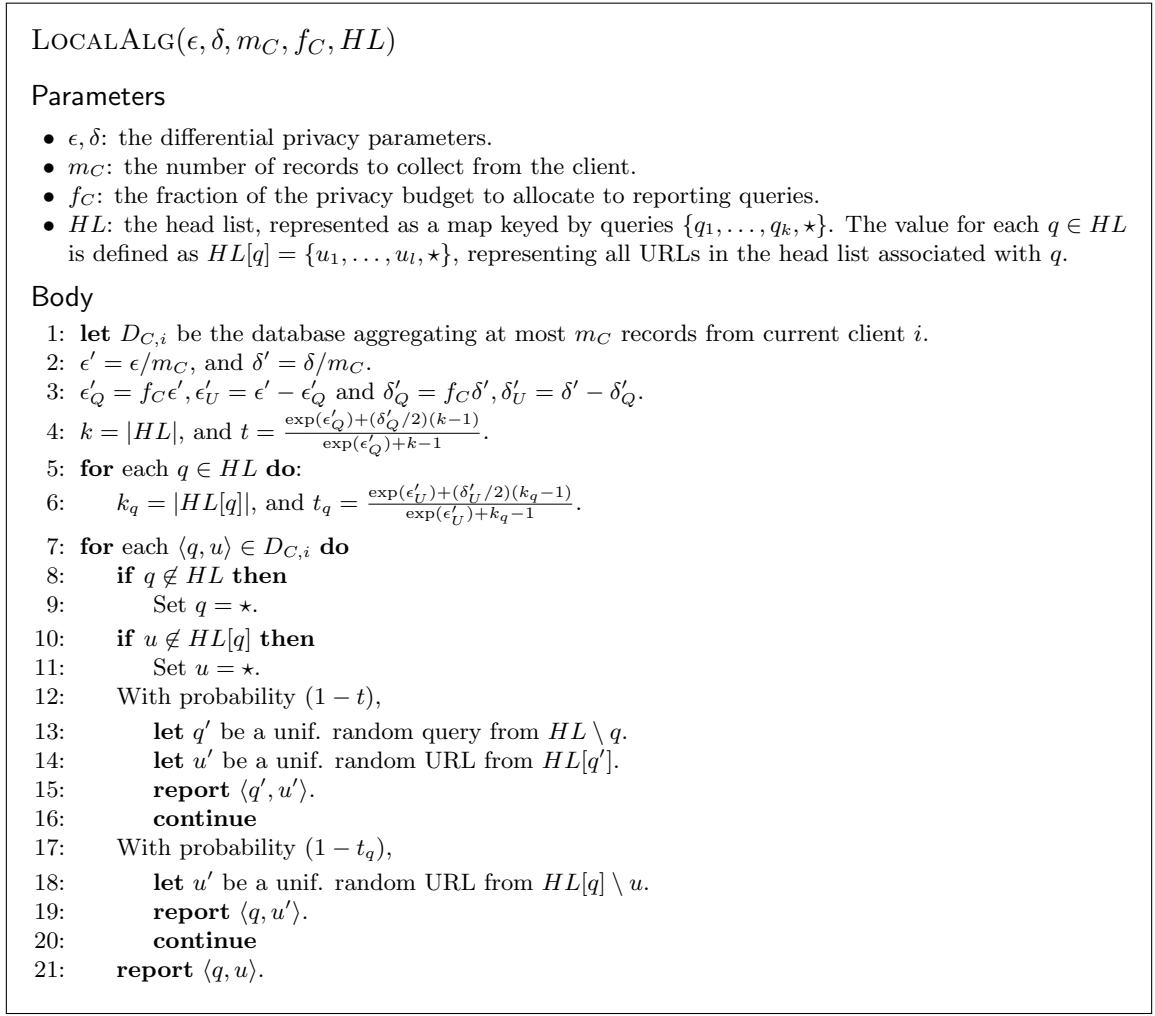

\raggedright	
	\begin{framed}		
		\begin{algorithmic}[1]{
			\params
                \footnotesize
				\begin{itemize}
					\item $\epsilon, \delta$: the differential privacy parameters.
					\item $m_C$: the number of records to collect from the client.
					\item $f_C$: the fraction of the privacy budget to allocate to reporting queries.
					\item $HL$: the head list, represented as a map keyed by queries $\{q_1,\dots,q_k,\star\}$. The value for each $q \in HL$ is defined as $HL[q] = \{u_1,\dots,u_l,\star\}$, representing all URLs in the head list associated with $q$.
				\end{itemize}
			}
			{\textsc{LocalAlg}$(\epsilon, \delta, m_C, f_C, HL)$}
			\Let $D_{C,i}$ be the database aggregating at most $m_C$ records from current client $i$.
			\State $\epsilon^\prime = \epsilon / m_C$, and $\delta^\prime = \delta / m_C$. \label{begin-define-vars}
			\State $\epsilon^\prime_Q = f_C\epsilon^\prime, \epsilon^\prime_U = \epsilon^\prime - \epsilon^\prime_Q$ and $\delta^\prime_Q = f_C\delta^\prime, \delta^\prime_U = \delta^\prime - \delta^\prime_Q.$
			\State $k = |HL|$, and $t = \frac{\exp(\epsilon^\prime_Q)+(\delta^\prime_Q / 2)(k-1)}{\exp(\epsilon^\prime_Q)+k-1}$. \label{end-define-vars}
			\For {each $q \in HL$}: 
				\State $k_q = |HL[q]|$, and $t_q = \frac{\exp(\epsilon^\prime_U)+ (\delta^\prime_U / 2)(k_q-1)}{\exp(\epsilon^\prime_U)+k_q-1}$.
			\EndFor
			\For {each $\langle q,u\rangle \in D_{C,i}$}\label{local-loop}
				\If {$q \not \in HL$}
					\State Set $q = \star$.
					\EndIf
				\If {$u \not \in HL[q]$}
					\State Set $u = \star$.
				\EndIf
				
				\State \textrm{With probability} $(1-t)$,
				\Indent
					\Let $q^\prime$ be a unif. random query from $HL \setminus q$.
					\Let $u^\prime$ be a unif. random URL from $HL[q^\prime]$.
					\State \textrm{\textbf{report} $\langle q^\prime,u^\prime\rangle$.} 
					\Continue
				\EndIndent
				
				\State \textrm{With probability} $(1-t_q)$,
				\Indent
					\Let $u^\prime$ be a unif. random URL from $HL[q] \setminus u$.
					\State \textrm{\textbf{report} $\langle q,u^\prime\rangle$.} 
					\Continue
				\EndIndent
				\State \textrm{\textbf{report} $\langle q,u\rangle$}.
				\EndFor
		\end{algorithmic}
	\end{framed}
	\caption{Algorithm executed by each client for privately reporting their records.}
	\label{fig:local-alg}
	\vspace{-0mm}
\end{figure}

\point{Algorithms for client data collection (Figures~\ref{fig:client-alg},~\ref{fig:local-alg})}
Figure~\ref{fig:client-alg} defines the algorithm for the client group.
Here, records are no longer treated as a single entity, but rather in a two-stage process: first privatizing the query, then privatizing the URL.
This helps optimize utility in the setting where the number of queries is significantly larger than the number of URLs associated with each query.
Privatization as achieved by following a generalization of the randomized response mechanism introduced by~\cite{warner1965randomized}, and utilizes the head list obtained from the opt-in group in order to perform the privatization locally by each client.
At its core, the privatization is achieved by reporting the true record with a certain bounded probability, and otherwise, randomizing the report among all the other records in the head list.

The fact that the head list (approximating the set of the most frequent records) is available to each client plays a crucial role in improving the utility of the data produced by this privatization algorithm compared to the previously known algorithms operating in the local privacy model.
This allows use of the entire privacy budget to report the true value, rather than having to allocate some of it for estimating an analogue of the head list, as done in~\cite{fanti2016building, qin2016heavy}.
Another distinction from the standard randomized response mechanism is our utilization of $\delta$.

Note that the choices of $m_C$ and $f_C$ are not related to privacy constraints, and can be chosen by \tool's developer to optimize utility goals, as will be discussed in Section~\ref{sec:parameter-choices}.

The local nature of the reporting, using a randomization procedure that can report any record with some probability, induces a predictable bias to the distribution of reported records.
To account for this, a denoising procedure must be performed in order to compute proper estimates.

These probability estimates, denoted $\hat{p}_C$, along with variance estimates are then passed to the \textsc{BlendProbabilities} part of \tool.
The technical discussion of the algorithm's privacy properties and variance estimate computations follow in Sections~\ref{sec:client-technical} and~\ref{sec:blending-technical}.\\

\begin{figure}[h!]
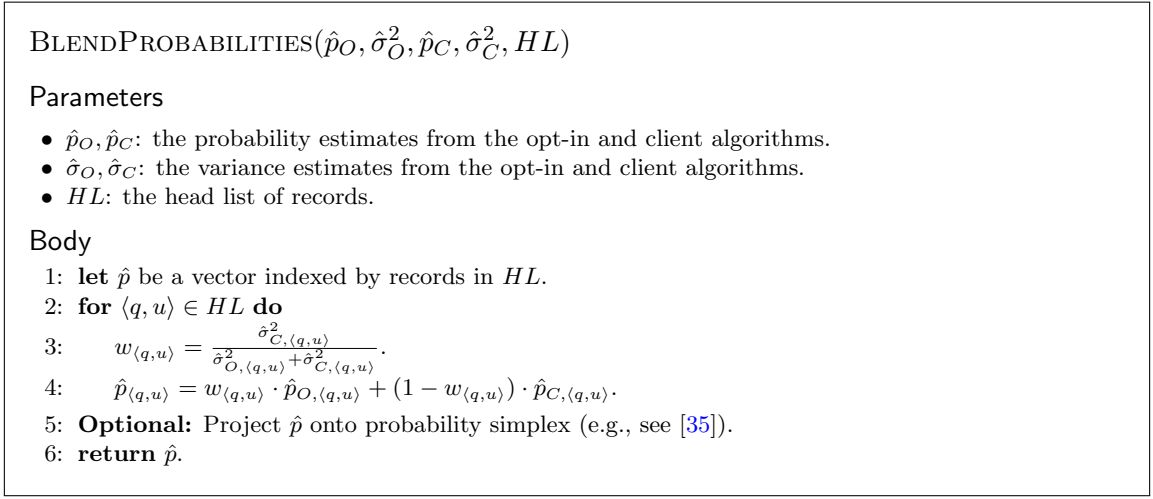

\raggedright 
	\begin{framed}		
		\begin{algorithmic}[1]{
			\params
                \footnotesize
				\begin{itemize}
					\item $\hat{p}_O, \hat{p}_C$: the probability estimates from the opt-in and client algorithms.
					\item $\hat{\sigma}_O, \hat{\sigma}_C$: the variance estimates from the opt-in and client algorithms.
					\item $HL$: the head list of records.
				\end{itemize}
			}
			{\textsc{BlendProbabilities}$(\hat{p}_O, \hat{\sigma}^2_O, \hat{p}_C, \hat{\sigma}^2_C, HL)$}
			\Let $\hat{p}$ be a vector indexed by records in $HL$.
			\For {$\langle q,u\rangle \in HL$}
				\State $w_{\langle q,u\rangle} = \frac{\hat{\sigma}^2_{C,\langle q,u\rangle}}{\hat{\sigma}^2_{O,\langle q,u\rangle} + \hat{\sigma}^2_{C,\langle q,u\rangle}}$.\label{line:blend-weight}
				\State $\hat{p}_{\langle q,u\rangle} = w_{\langle q,u\rangle}\cdot\hat{p}_{O,\langle q,u\rangle} + 
						(1-w_{\langle q,u\rangle})\cdot\hat{p}_{C,\langle q,u\rangle}.$
			\EndFor
			\State \textbf{Optional:} Project $\hat{p}$ onto probability simplex (e.g., see~\cite{wang2013projection}).
			\State \Return $\hat{p}$.
		\end{algorithmic}
	\end{framed}
	\caption{Algorithm for combining record probability estimates from opt-in and client estimates.}
	\label{fig:blending-alg}
\end{figure}

\point{Algorithm for Blending (Figure~\ref{fig:blending-alg})}
The blending portion of the algorithm combines the estimates produced by the opt-in and client probability-estimation algorithms by taking into account the sizes of the groups and the amount of noise each algorithm respectively added.
This produces a blended probability estimates $\hat{p}$ which, in expectation, is more accurate than either group produced individually.
The procedure for blending is not subject to privacy constraints, as it operates on the data whose privacy has already been ensured by previous steps of \tool.
The motivation and technical discussion of this algorithm follows in Section~\ref{sec:blending-technical}.\\

\section{Technical Details}
\label{sec:technical}
We now present further technical details related to the instantiations of the sub-algorithms for \tool, such as privacy proofs and the motivation for \textproc{BlendProbabilities}.\\

\subsection{Opt-in Data Algorithms}\label{sec:opt-in-technical}
Differential privacy of the algorithms handling opt-in client data follows directly from previous work.

\begin{theorem}(\cite{korolova2012protecting})
	\textproc{CreateHeadList} guarantees $(\epsilon, \delta)$-differential privacy if $m_O = 1, \epsilon > \ln(2),$ and $\tau \geq 1$.
\end{theorem}

\begin{theorem}(\cite{dwork2006calibrating})
	\textproc{EstimateOptinProbabilities} guarantees $(\epsilon, 0)$-differential privacy if $m_O = 1$.
\end{theorem}

\subsection{Client Data Algorithms}
\label{sec:client-technical}

\textsc{LocalAlg} is responsible for the privacy-preserving perturbation of each client's data before it gets sent to the server and \textsc{EstimateClientProbabilities} is responsible for aggregating the received privatized data into a meaningful statistic. We prove the privacy and explain the logic behind the aggregation procedure next.

\begin{theorem}
\label{theorem:localalg-priv}
	\textproc{LocalAlg} is $(\epsilon, \delta)$-differentially private.
\end{theorem}

\begin{proof}
See Appendix.
\end{proof}

The reports aggregated by the client algorithm form an empirical distribution over the records (and queries).
This distribution is biased in an explicit way, as described by the reporting process, creating a significantly flatter distribution relative to the true underlying distribution.
Since the noise addition process is known, the bias is also known, and can be used to ``unflatten'' the distribution as a post-processing step to obtain a more useful, unbiased estimate of the record distribution.
We refer to this as ``denoising'' the reported empirical distribution $\hat{r}_C$ to obtain the final estimate from the client algorithm, $\hat{p}_C$.

\vspace{0.3cm}
\begin{observation}
\label{observation:client-denoising}
$\hat{p}_C$ gives the unbiased estimate of record and query probabilities under \textproc{EstimateClientProbabilities}.
\end{observation}

\begin{proof}
See Appendix.
\end{proof}

\subsection{Blending}
\label{sec:blending-technical}

The opt-in algorithm and client algorithm both output independent estimates $\hat{p}_O$ and $\hat{p}_C$ of the record distribution $p$.
The question we address now is how to best combine these estimates using the information available.

A standard way to measure the quality of an estimate is by its variance.
Although it may seem natural to choose the estimate with lower variance as the final estimate $\hat{p}$, it is possible to achieve a better estimate by jointly utilizing the information provided by both algorithms.
This is because the errors in these estimates come from different sources.
The error in the estimates obtained from the opt-in algorithm is due to the addition of Laplace noise, whereas the error in the estimates obtained from the client algorithm is due to randomizing the true record over the set of records in the head list.
Thus, our goal is to determine the variances of the estimates obtained from the two algorithms and use these variances to \textit{blend} the estimates in the best way.

More formally, for each record $\langle q,u\rangle$ let $\sigma^2_{O,\langle q,u\rangle}$ and $\sigma^2_{C,\langle q,u\rangle}$ be the variances of the opt-in and client algorithms estimates of $\hat{p}_{O, \langle q,u\rangle}$ and $\hat{p}_{C,\langle q,u\rangle}$ respectively.
Since these variances depend on the underlying distribution, which is unknown a priori, we will compute sample variances $\hat{\sigma}^2_{O,\langle q,u\rangle}$ and $\hat{\sigma}^2_{O,\langle q,u\rangle}$ instead.
For each record $\langle q,u\rangle$, we will weight the estimate from the opt-algorithm by $w_{\langle q,u\rangle}$ and the estimate from the client algorithm by $(1-w_{\langle q,u\rangle})$, where $w_{\langle q,u\rangle}$ is defined as in line~\ref{line:blend-weight} of \textproc{BlendProbabilities}.
The optional step of projecting the blended estimates (e.g., as in \cite{wang2013projection}) ensures that the estimates sum to 1 and are non-negative.

Theorem~\ref{theorem:opt-in-variance} presents our computation of the sample variance of \textproc{EstimateOptinProbabilities}, Theorem~\ref{theorem:client-variance} presents our computation of the sample variance of \textproc{EstimateClientProbabilities}, and Theorem~\ref{theorem:weighting-scheme} motivates the weighting scheme used in \textproc{BlendProbabilities}.

For the variance derivations, we make an explicit assumption that each piece of reported data is drawn independently and identically from the same underlying distribution.
This is reasonable when comparing data across users. By setting $m_O = m_C = 1$, we remove the need to assume iid data \textit{within} each user's own data, while simplifying our variance computations.
We show in Section~\ref{sec:expt} that \tool achieves high utility even when $m_O = m_C = 1$.\\

\begin{theorem}
\label{theorem:opt-in-variance}
If $m_O = 1$ then the unbiased variance estimate for the opt-in group's record probabilities can be computed as:
$\hat{\sigma}^2_{O,\langle q,u\rangle} = \frac{|D_T|}{|D_T|-1}\left(\frac{\hat{p}_{O,\langle q,u\rangle}(1-\hat{p}_{O,\langle q,u\rangle})}{|D_T|} + 2 \left(\frac{b_T}{|D_T|}\right)^2\right).$
\end{theorem}

\begin{proof}
See Appendix.
\end{proof}

Note that in line~15 of \textproc{EstimateOptinProbabilities}, the use of this sample variance expression in re-computing $\hat{\sigma}^2_{O,\langle \star, \star \rangle}$ is not statistically valid, so our computation of $\hat{p}_{O,\langle \star, \star \rangle}$ and $\hat{p}_{\langle \star, \star \rangle}$ is sub-optimal. Despite that, our overall utility, which does not include $\star$, is good (see Section~\ref{sec:expt}).\\

\vspace{0.3cm}
\begin{theorem}
\label{theorem:client-variance}
If $m_C = 1$ then the unbiased variance estimate for the client group's record probabilities can be computed as:
\footnotesize
\begin{align*}
\hat{\sigma}^2_{C,\langle q,u\rangle}& = \frac{|D_C|}{t^2\bigl(t_q - \frac{1-t_q}{k_q-1}\bigr)^2 (|D_C|-1)} \cdot\\
		&\Bigl( \frac{\hat{r}_{C,\langle q,u\rangle}(1-\hat{r}_{C,\langle q,u\rangle})}{|D_C|}
			\quad + \bigl(\frac{1-t}{(k-1)k_q} - t\frac{1-t_q}{k_q-1} \bigr)^2 \hat{\sigma}^2_{C,q} + 2 
			\bigl(\frac{1-t}{(k-1)k_q} -~t\frac{1-t_q}{k_q-1} \bigr)\frac{\hat{r}_{C,\langle q,u\rangle}(1-\hat{r}_{C,q})}{|D_C| (t-\frac{1-t}{k-1})} \Bigr).
\end{align*}
\end{theorem}
\normalsize

\begin{proof}
See Appendix.
\end{proof}

\vspace{0.3cm}
\begin{theorem}
\label{theorem:weighting-scheme}
If $\hat{\sigma}^2_{O,\langle q,u\rangle}$ and $\hat{\sigma}^2_{C,\langle q,u\rangle}$ are sample variances of $\hat{p}_{O,\langle q,u\rangle}$ and $\hat{p}_{C, \langle q,u\rangle}$ respectively, and the blended estimate is the convex combination $\hat{p}_{\langle q,u\rangle} = w_{\langle q,u\rangle}\cdot\hat{p}_{O,\langle q,u\rangle} + (1-w_{\langle q,u\rangle})\cdot\hat{p}_{C,\langle q,u\rangle}$, then the sample variance optimal weighting is given by $w_{\langle q,u\rangle} = \frac{\hat{\sigma}^2_{C,\langle q,u\rangle}}{\hat{\sigma}^2_{O,\langle q,u\rangle} + \hat{\sigma}^2_{C,\langle q,u\rangle}}$.
\end{theorem}
\begin{proof}
See Appendix.
\end{proof}

\subsection{Discussion}
\label{sec:discussion-technical}
We have intentionally used (slight modifications) of existing algorithms for \tool's sub-algorithms in order to demonstrate the power of the blended approach within the hybrid privacy model.
However, it is conceivable (see Section~\ref{sec:related}) that the sub-algorithms themselves can be improved, yielding further improvements in the utility achieved by \tool.

Operating in the hybrid model is most beneficial utility-wise if the opt-in user records and client user records come from the same distribution -- i.e., the two groups have fairly similar observed search behavior.
If that is not the case, the differential privacy guarantees still hold, but the accuracy of \tool 's estimates may decrease.\\

\section{Experimental Evaluation}
\label{sec:expt}

\subsection{Utility Metrics}
\label{sec:expt-utility}
One pitfall in much of the research in the area of differential privacy is an insufficient emphasis on the utility loss due to privacy constraints.
We designed \tool with an eye toward preserving the utility of the eventual results in the two applications we explore in this paper: trend computation and local search, as described in Section~\ref{sec:intro}.
We use two domain-specific utility metrics to assess the loss of utility, the L1 metric and NDCG. \\

\point{L1}
L1 is the Manhattan distance between the estimate and actual probability vectors, in other words,
$L1=\sum_{i}{|\hat{p}_i-p_i|}.$ The smaller the L1, the better.\\

\point{NDCG}
NDCG is a standard measure of search quality~\cite{jarvelin2002cumulated,valizadegan2009learning} that takes into account the order of queries by performing \emph{discounting}.
In particular, most popular queries at the \emph{head} of the search have a higher weight, whereas the relative significance of the less popular queries is reduced. 
The relevance, or gain, of an item at position $i$ in the ranked list is measured using a graded relevance score defined as
$\mathit{rel}_i = \frac{n_i}{\sum_{j} n_j},$
where $n_j$ is the number of occurrences of the item in position $j$ in the given dataset.
The closer $i$'s estimated rank is to its true rank, the larger the gain. For a \emph{head} of $k$ top elements, the estimated rank list is computed as 
$DCG_k={\sum^k_{i=1}\frac{2^{rel_{i}}-1}{\log_2(i+1)}}.$

Here, the discounting happens because of the $\log_2(i)$ factor that diminishes the effect of later queries.
This value is normalized by the Ideal DCG~($IDCG_k$), in which the estimated and the actual ranking are exactly the same, to obtain a value that ranges between~$0$ and~$1$. 

Since we operate on records rather than just queries, we utilize a generalization of the traditional NDCG score.
Here, we compute the NDCG of each query's URL list, $NDCG^q$, as specified above, and then compute the DCG of the queries as
$DCG^Q_k={\sum^k_{i=1}\frac{2^{rel_{i}}-1}{\log_2(i+1)} \cdot NDCG^i}.$

The final NDCG computation is then $DCG^Q_k$ normalized by the analogous Ideal DCG ($IDCG^Q_k$).
In a way, our computation considers an NDCG of NDCGs, which makes it even harder for us to maintain consistently high NDCG values when compared to the query-only case. This formulation takes the \emph{ranking} and frequencies from the dataset into account, not the actual score that our algorithm outputs. Since changes to the score may not result in ranking changes, L1 is an even less forgiving measure than NDCG.

Since the purpose of \tool is to estimate probabilities of the top records, we discard the artificially added $\star$ queries and URLs and rescale $rel_i$ prior to L1 and NDCG computations. However, since we use the method of~\cite{wang2013projection} in \textsc{BlendProbabilities}, the probability estimates involving $\star$ have a minor implicit effect on the L1 and NDCG scores.\\

\subsection{Experimental Setup}
For our experiments, we use the AOL search logs, released in~2006 and the Yandex search dataset\footnote{\url{https://www.kaggle.com/c/yandex-personalized-web-search-challenge/data}}, released in~2013. Figure~\ref{fig:dataset-stats} describes their characteristics.

\begin{figure}[h!]
\centering
\setlength{\tabcolsep}{16pt}
\begin{tabular}{lrr}
& \bf AOL & \bf Yandex \\
\midrule
    Dataset on disk & {1.75}~GB  & {16}~GB\\
    Unique queries    &  \nnum{4811646} & \nnum{13171961}\\
    Unique clients    & \nnum{519371}   & \nnum{4970073}\\
    Unique URLs       & \nnum{1620064}  &  \nnum{12702350}\\
\bottomrule    \\
\end{tabular}
\vspace{-5mm}
\caption{Dataset statistics.}
\label{fig:dataset-stats}
\end{figure}

\point{Data analysis} To familiarize the reader with the approach we used for assessing result quality, Figure~\ref{fig:frequencies} shows the top-10 most frequent queries in the AOL dataset, with the estimates given by the different ``ingredients'' of \tool.

The table is sorted by column~2, which contains the non-private, empirical probabilities from the AOL dataset with 1 random record sampled from each user.
Column~3 contains the final query probability estimates outputted by \tool.
Each algorithm computes probability estimates over the records in the head list; to obtain query probability estimates from these, we simply aggregate the probabilities associated with each URL for a given query (columns~4 and~6). The sample variance of these aggregated probabilities, used for blending, is na\"ively computed as in Theorem~\ref{theorem:opt-in-variance}.
Column~5 is the \textproc{EstimateClientProbabilities}' estimate of the query probabilities, since it directly computes these values.
While column~6 is not used for blending in trend computation (where only query probability estimates are produced), columns~4,~5, and~6 are used by the full \tool algorithm when it comes to blending entire records.
Regressions, i.e., estimates that appear out of order relative to column~2, are shown in red.

\newcolumntype{L}[1]{>{\raggedright\let\newline\\\arraybackslash\hspace{0pt}}m{#1}}
\newcolumntype{C}[1]{>{\centering\let\newline\\\arraybackslash\hspace{0pt}}m{#1}}
\newcolumntype{R}[1]{>{\raggedleft\let\newline\\\arraybackslash\hspace{0pt}}m{#1}}
\newcommand{\regression}[1]{\color{red}\textbf{#1}}

\begin{figure}[h]
\newcommand{\scaling}{.95}
\setlength{\tabcolsep}{4pt}
\centering
\renewcommand{\arraystretch}{1.0}
    \begin{tabular}{|l|c|c|c|c|c|}
    \hline
    & \bf AOL & \textbf{\tool} & \bf Opt-in & \bf Client & \bf Client \\
    \textbf{Query} & 
    \multicolumn{1}{c|}{{\textbf{dataset}}} & 
    \multicolumn{1}{c|}{\textbf{estimate}} & 
    \multicolumn{1}{c|}{\textbf{estimate}} & 
    \multicolumn{1}{c|}{\textbf{estimate}} & 
    \multicolumn{1}{c|}{\textbf{estimate}} \\
    \multicolumn{1}{|c|}{{}} &
    \multicolumn{1}{c|}{{\textbf{prob}}} & 
    \multicolumn{1}{c|}{$\hat{p}_q$} & 
    \multicolumn{1}{c|}{$\sum_u{\hat{p}_{O,\langle q,u\rangle}}$} & 
    \multicolumn{1}{c|}{{$\hat{p}_{C,q}$}} & 
    \multicolumn{1}{c|}{$\sum_u{\hat{p}_{C,\langle q,u\rangle}}$} \bigstrut\\
    \hline
	$\star$        & 0.9121 & 0.9144 & 0.9148 & 0.9143 & 0.9143 \bigstrut[t]\\\hline
	google         & 0.0211 & 0.0211 & 0.0220 & 0.0210 & 0.0210 \\\hline
	yahoo     	   & 0.0067 & 0.0081 & 0.0061 & 0.0088 & 0.0088 \\\hline
	google.com     & 0.0066 & 0.0075 & \regression{0.0083} & 0.0073 & 0.0073 \\\hline
	myspace.com    & 0.0055 & 0.0046 & 0.0034 & 0.0052 & 0.0052 \\\hline
	mapquest       & 0.0055 & \regression{0.0062} & \regression{0.0051} & \regression{0.0066} & \regression{0.0066} \\\hline
	yahoo.com      & 0.0048 & 0.0047 & \regression{0.0057} & 0.0043 & 0.0043 \\\hline
	www.google.com & 0.0044 & 0.0038 & 0.0043 & 0.0035 & 0.0035 \\\hline
	myspace        & 0.0034 & 0.0030 & 0.0031 & 0.0030 & 0.0030 \\\hline
	ebay  		   & 0.0030 & 0.0030 & 0.0030 & 0.0029 & 0.0029 \\\hline
    \end{tabular}

  \caption{Comparison of probability estimates for top-10 most popular AOL queries. Parameter choices are shown in Figure~\ref{table:parameters} (except with $\epsilon=3$ used here).}
  \label{fig:frequencies}
\end{figure}

The biggest takeaway is that the numbers in columns~2 and~3 are similar to each other, with \tool's usage resulting in only one regression.

\tool compensates for the weaknesses of both the opt-in and the client estimates.
Specifically, despite the opt-in group having several regressions in this particular instance, combining the opt-in and client-data compensates for that, resulting in only \empirical{one} regression.

\subsection{Experimental Results}
We formulate questions for our evaluation as follows: how are \tool's parameters chosen (Section~\ref{sec:parameter-choices}), how does \tool perform compared to alternatives (Section~\ref{sec:ccs}), and how robust are our findings (Section~\ref{sec:robustness})?

\subsubsection{Algorithmic and Parameter Choices}\label{sec:parameter-choices}
\tool has a handful of parameters, some of which can be viewed as given externally (by the laws of nature, so to speak), and others whose choice is purely up to the entity deploying \tool. We now describe and, whenever possible, motivate, our choices for these.\\

\point{Privacy parameters, $\epsilon$ and $\delta$}
We view $\epsilon$ and $\delta$ as privacy parameters given to us externally (e.g., by what is a common practice for differentially private algorithms in the industry~\cite{appleepsilon, Apple, erlingsson2014rappor}).
We use a $\delta$ that is larger for the AOL dataset than for the Yandex dataset to reflect that the Yandex dataset contains data of more users.

We use the same $\epsilon$ and $\delta$ values for the opt-in and client users. 
From a behavioral perspective, this reduces a user's opt-in decision down to one purely of trust towards the curator.\\

\point{Opt-in and client group sizes, $|O|$ and $|C|$}
The relative sizes of opt-in group and client group, $|O|$ and $|C|$, respectively, can be viewed as exogenous variables which are dictated by the trust that users place in the search engine\footnote{In the future, as differential privacy gains widespread adoption, it is conceivable that the values of the privacy parameters may affect their relative sizes; for example, the smaller the $\epsilon$, the more users are willing to ``opt-in". However, this relationship is out of the scope of this work.}.
We choose \empirical{$5\%$} for AOL and \empirical{$3\%$} for Yandex for the fraction of opt-in users because they seem reasonable while still allowing us to effectively demonstrate the utility benefits of \tool.\\

\point{The number of records to collect from each opt-in user, $m_O =1$}
This is a choice necessitated by the privacy constraints of the \textsc{CreateHeadList} algorithm.\\

\noindent The choices for remaining parameters: $m_C, f_C, f_O, M$ are driven purely by utility considerations.\\

\point{The number of records to collect from each client, $m_C =1$}
Across a range of experimental values, collecting 1 record per user always yielded greatest utility, justifying this parameter choice. Two factors account for this: 1) the privacy budget must be split across a client's reports, and 2) the accuracy of our algorithm relies on uncorrelated reports, an assumption which likely does not hold in practice within a given user's set of records.\\

\begin{figure}[tb]
\begin{subfigure}{0.49\columnwidth}
\centering
\ifpdf
    \includegraphics[width=.99\columnwidth]{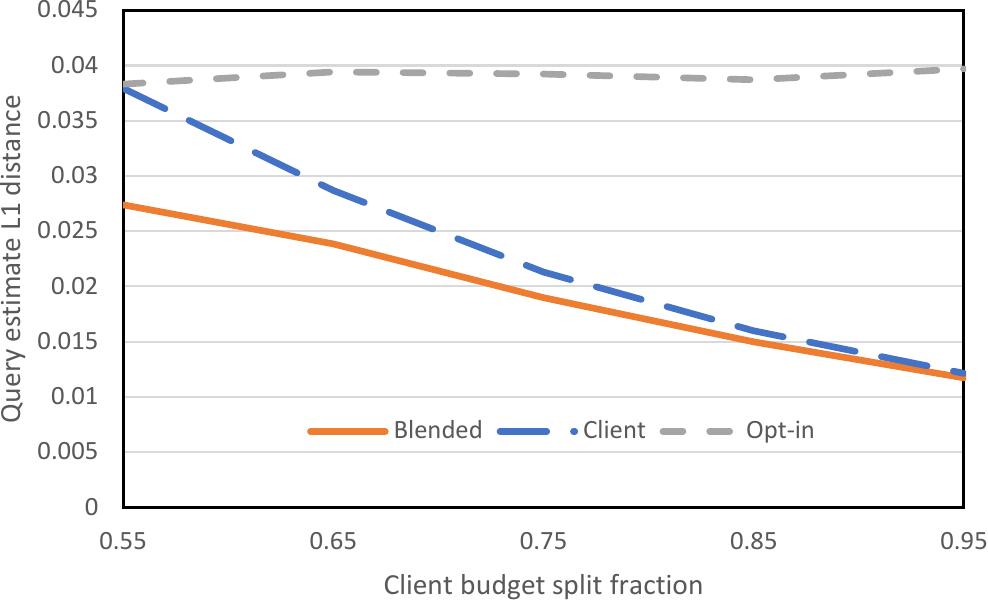}
\fi
\caption{L1}
\end{subfigure}
\begin{subfigure}{0.49\columnwidth}
\centering
\ifpdf
    \includegraphics[width=.99\columnwidth]{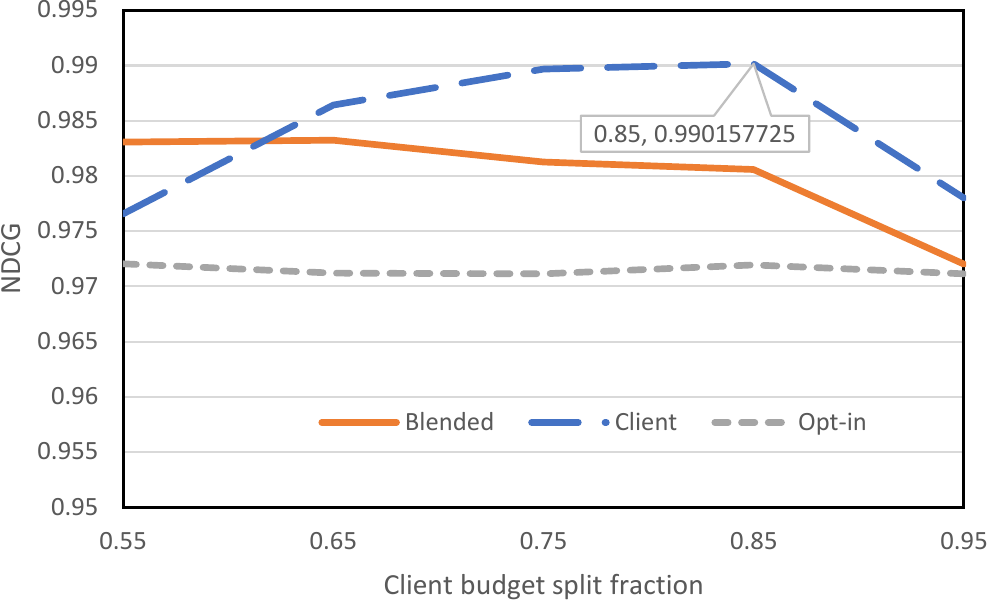}
\fi
\caption{NDCG}
\end{subfigure}
\caption{Comparing AOL dataset results across a range of budget splits for client, opt-in, and blended results.}
\label{fig:blended-vs-not}
\end{figure}

\point{How to split the privacy budget between query and URL reporting for clients, $f_C = 0.85$}
Figure~\ref{fig:blended-vs-not} shows the effects of the budget split on both the L1 and NDCG metrics. 
Unsurprisingly, Figure~\ref{fig:blended-vs-not}a shows that the larger the fraction of client algorithm's budget dedicated to query estimation as opposed to URL estimation, the better the L1 score for the client and \tool results. 
The NDCG metric in Figure~\ref{fig:blended-vs-not}b shows a trade-off that emerges as we assign more budget to the queries, de-emphasizing the URLs. The client algorithm NDCG value peaks at a budget split of~\empirical{0.85}; choosing a split above this point induces a significant drop in the blended NDCG values.
Note that the grey opt-in line in Figure~\ref{fig:blended-vs-not}b is constant, as the opt-in group is not affected by the budget split.\\

\point{What fraction of opt-in data to use for creating the headlist, $f_O = 0.95$}
We choose $f_O = 0.95$ because our goal is to build a large candidate head list, and unless we allocate most of the opt-in user data to building such a head list (algorithm \textsc{CreateHeadList}), our subsequent results may be accurate but apply only to a small number of records, whereas in order to speak of an effective local search application in practice we need to amass a headlist of at least double or triple digits in size\tr{~\cite{??}}.
Even without looking at experimental data, this choice makes sense: our opt-in group size is small relative to our client group size, and it is difficult to generate a head list in the local privacy model -- so it makes sense to utilize most of the opt-in group's data for the task that is most difficult in the local model.\\

\point{What should be the final size of the set for which we provide probability estimates, $M$}
The choice of $M$ is influenced by competing considerations.
The larger the head list for which we provide the probability estimates, the more effective the local search application (subject to those probability estimates being accurate).
However, as desired head list size increases, the accuracy of our estimates drops (most notably due to client data privatization). We want to strike a balance that allows us to get a sensibly large record set with reasonably accurate probability estimates for it. 
We choose $M=50$ and $M=500$ for the AOL and Yandex datasets, to reflect their differing sizes.\\

\begin{wrapfigure}{r}{5cm}
	\vspace{-4mm}
	\centering
	\setlength{\tabcolsep}{1pt}
	\begin{tabular}{ |c|rR{15mm}| }
		\hline
		\bf Parameter & \bf AOL & \bf Yandex \\
		\hline
		$\epsilon$ & 4 & 4 \\
		$\delta$ & $10^{-5}$ & $10^{-7}$ \\
		$\frac{|O|}{|O| + |C|}$ & 5\% & 3\% \\
		$m_O$ & 1 & 1 \\
		$m_C$ & 1 & 1\\
		$f_O$ & 0.95 & 0.95 \\
		$f_C$ & 0.85 & 0.85 \\
		$M$ & 50 & 500 \\
		\hline
	\end{tabular}
	\caption{Default parameters used in experiments.}
	\label{table:parameters}
	\vspace{-7mm}
\end{wrapfigure}

\noindent Subsequently, we use the parameters shown in Figure~\ref{table:parameters}, unless explicitly stated.

\subsubsection{Utility Comparison to Alternatives}\label{sec:ccs}
\begin{figure}[tb]
\centering
\ifpdf
    \includegraphics[width=.55\columnwidth]{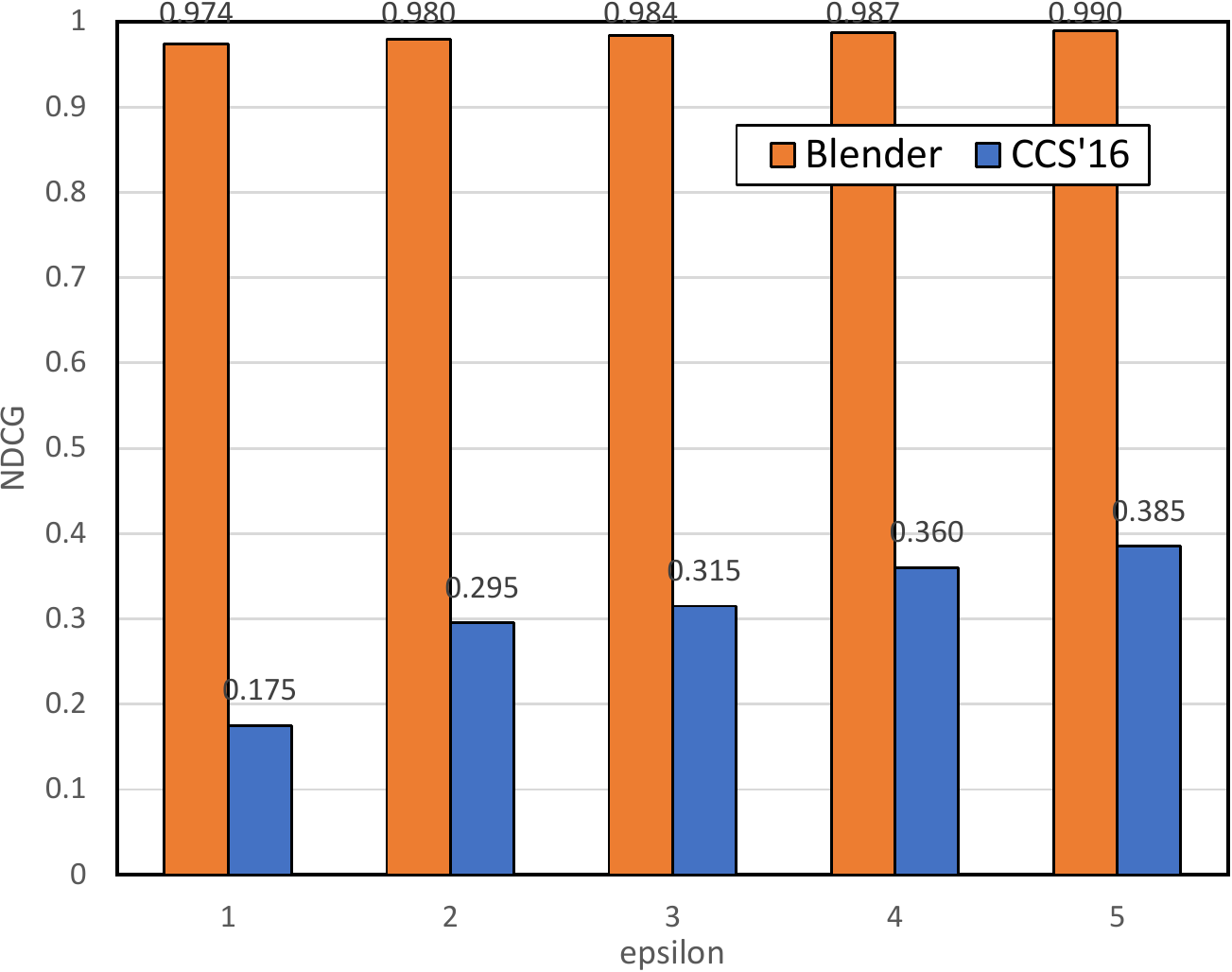}
\fi
\caption{Comparing to the results in the CCS'16 paper by Qin~\etal across a range of $\epsilon$ values; head list size=10.}
\label{fig:ccs}
\end{figure}

The closest related work is a recent paper by Qin~\etal~\cite{qin2016heavy} in which they provide a utility evaluation of their state-of-the-art algorithm on the AOL dataset for the headlist size of~10.
Given the NDCG data that they make available in the paper, we perform a direct comparison with \tool across $\epsilon$ values.
We plot the outcome of the comparison in Figure~\ref{fig:ccs}, which shows the NDCG values achieved by \tool and by Qin~\etal~\cite{qin2016heavy} for $\epsilon$ values between~1--5.
Across the entire range of the privacy parameter, our NDCG values are above~\empirical{95\%}, whereas the reported NDCG values for Qin~\etal are in the~30\% range, at best.
We believe that given the intense focus on search optimization in the field of information retrieval, NDCG values as low as those of Qin~\etal are generally unusable.
Overall, \tool significantly outperforms what we believe to be the closest related research project. 

Qin~\etal and this work use different scoring functions.
Qin~\etal use a relevance score based purely on the rank of queries in the original AOL dataset; this results in penalizing misranked queries regardless of their underlying probabilities.
\tool's relevance scoring only relies on the underlying probabilities, so misranked items with similar underlying probabilities only have a small negative impact on the overall NDCG score; we believe this choice is justified. While it yields increased NDCG scores, \tool operates on records (rather than queries, as Qin~\etal does).
Because of this, the ``NDCG of NDCGs'' score used to evaluate \tool (Section~\ref{sec:expt-utility}) is a strictly less forgiving metric than the traditional NDCG score.
Thus, although simultaneously compensating for both differences would yield the ideal comparison, the comparison in Figure~\ref{fig:ccs} is reasonable.

\newcolumntype{L}[1]{>{\raggedright\let\newline\\\arraybackslash\hspace{0pt}}m{#1}}
\newcolumntype{C}[1]{>{\centering\let\newline\\\arraybackslash\hspace{0pt}}m{#1}}
\newcolumntype{R}[1]{>{\raggedleft\let\newline\\\arraybackslash\hspace{0pt}}m{#1}}

\subsubsection{Robustness}\label{sec:robustness}
We now discuss how the size of the opt-in group and the choice of the $\epsilon$ privacy parameter affect \tool's utility.\\
\begin{figure}[tb]
\begin{subfigure}{0.49\columnwidth}
\centering
\ifpdf
    \includegraphics[width=.99\columnwidth]{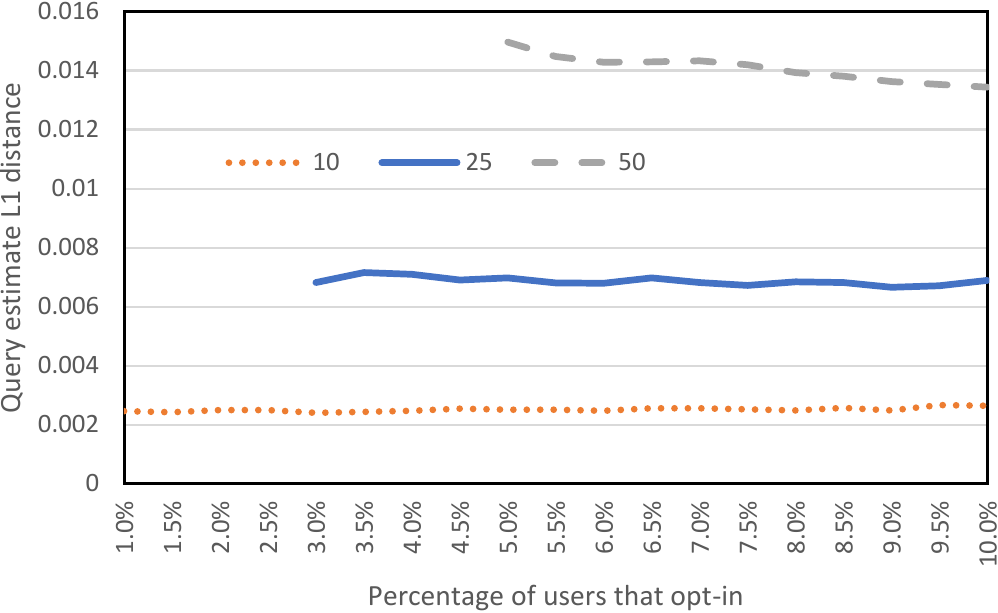}
\fi
\caption{AOL}
\end{subfigure}
\begin{subfigure}{0.49\columnwidth}
	\centering
	\ifpdf
	\includegraphics[width=.99\columnwidth]{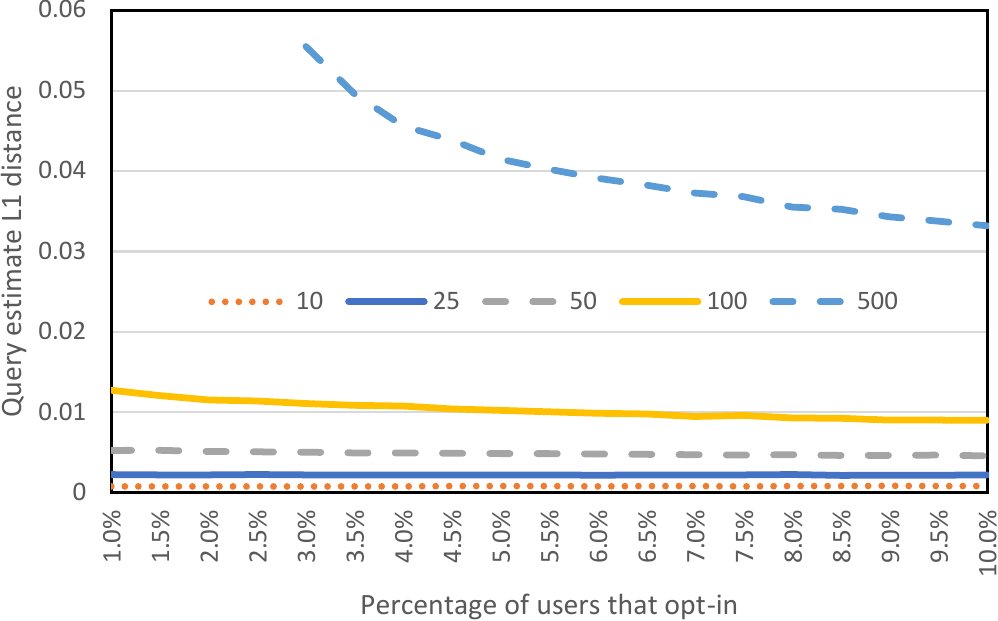}
	\fi
	\caption{Yandex}
\end{subfigure}
\caption{L1 as a function of the opt-in percentage.}
\label{fig:L1-by-optin-percentage}
\end{figure}

\point{Evaluation of trend computation}
Figure~\ref{fig:L1-by-optin-percentage}\footnote{Portions of lines do not appear on figures if the desired head list size was not reached (e.g., in Figure~\ref{fig:L1-by-optin-percentage}a, the line representing results for a head list of size 50 does not begin until 5\% because a head list of size 50 could not be generated with a lower opt-in percentage).} shows the L1 values as a function of the opt-in percentage ranging between~1\% and~10\%.
\begin{figure}[tb]
\centering
\begin{subfigure}{0.49\columnwidth}
\ifpdf
    \includegraphics[width=.99\columnwidth]{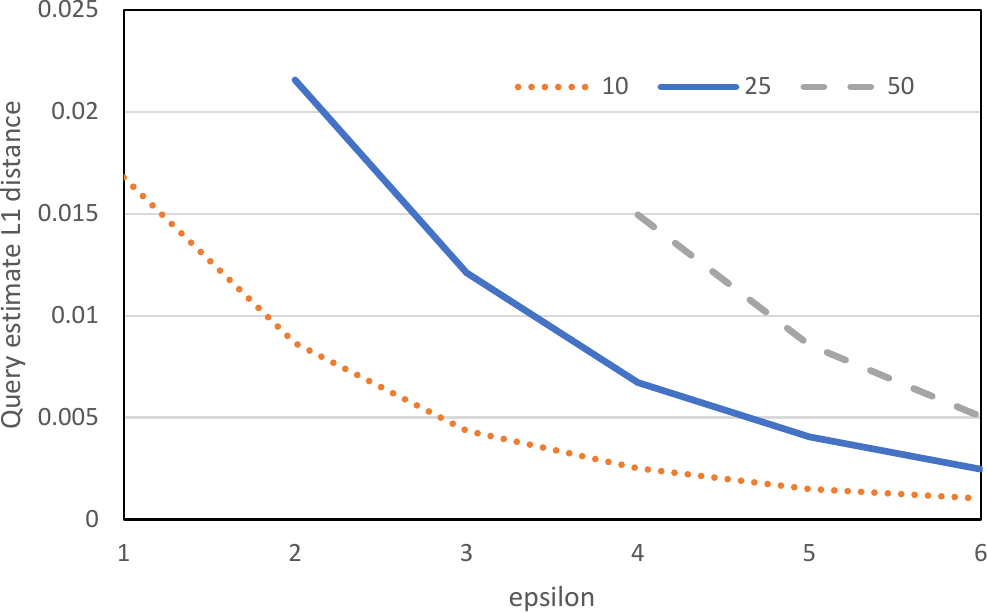}
\fi
\caption{AOL}
\label{fig:L1}
\end{subfigure}
\begin{subfigure}{0.49\columnwidth}
\ifpdf
    \includegraphics[width=.99\columnwidth]{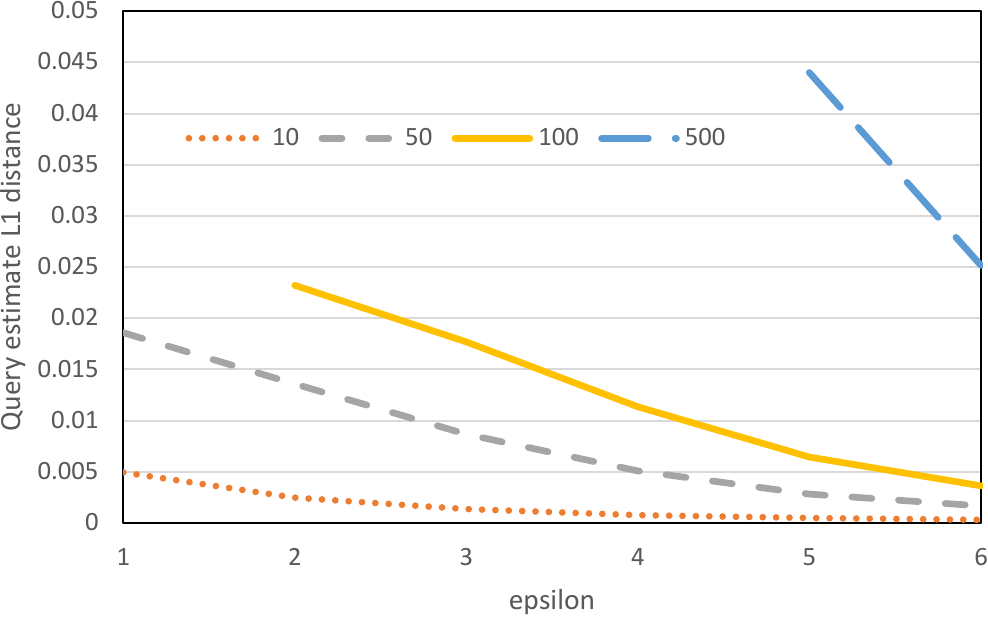}
\fi
\caption{Yandex}
\end{subfigure}
\caption{L1 statistics for AOL and Yandex datasets for various head list sizes and a range of $\epsilon$ values.}
\label{fig:L1-datasets}
\end{figure}

We believe that requiring opt-in percentages in excess of~10\% is likely to put undue strain on the system in terms of recruitment; simply put, finding enough opt-in users may provide difficult or impossible in the long run.
We see slight differences in the two datasets and across the various head list sizes. 
Some of the differences might be due to the fact that given the relatively small size of the AOL dataset, we need to consider higher opt-in percentages to get reasonably sized head lists and L1 values.
In fact, when we increase the opt-in percentage to~10\% for the AOL dataset, we see a slight decline in L1 values for the largest head list size similar to what is observed in Figure~\ref{fig:L1-by-optin-percentage}b for the Yandex dataset.
If our goal is to have head lists of~$500+$, we see that with the larger Yandex dataset, an opt-in percentage as small as~\empirical{3\%} is sufficient.
The main take-away from this is that when the opt-in group is large enough to attain the desired head list size, the trend computation results generally will be high quality in terms of the L1 values.

Figure~\ref{fig:L1-datasets} shows the L1 values as a function of $\epsilon,$ ranging from~1 to~6. 
For both datasets, we see a steady decline in the L1 metric, despite aggregating L1 values over longer estimate vectors. 
With more data in the Yandex dataset, we are able to hit small values of L1 (under \empirical{0.1}) with $\epsilon\ge1.$\\

\begin{figure}[tb]
\begin{subfigure}{0.49\columnwidth}
\centering
\ifpdf
    \includegraphics[width=.99\columnwidth]{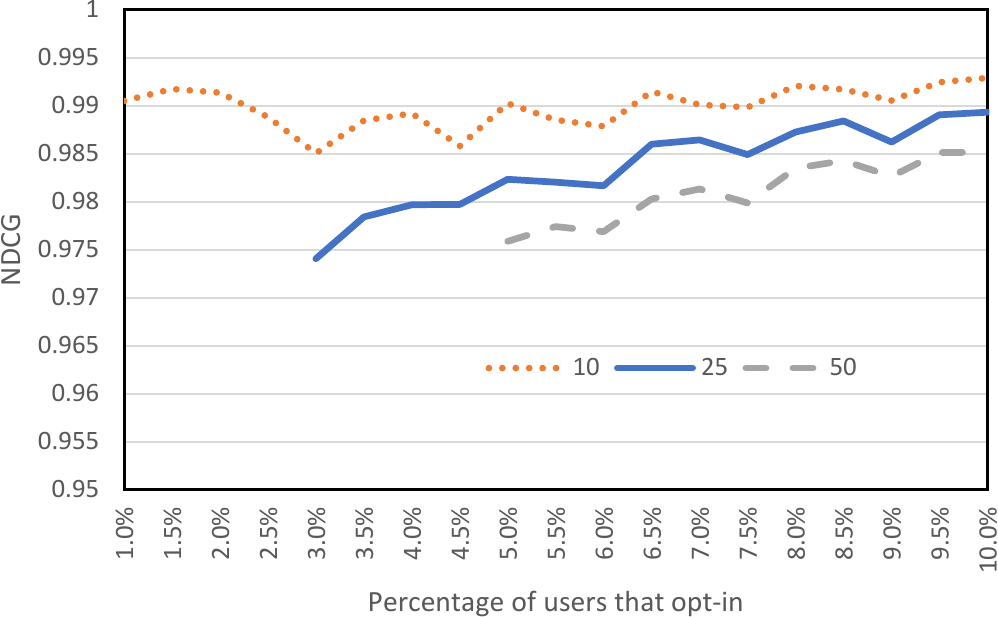}
\fi
\caption{AOL}
\end{subfigure}
\begin{subfigure}{0.49\columnwidth}
	\centering
	\ifpdf
	\includegraphics[width=.99\columnwidth]{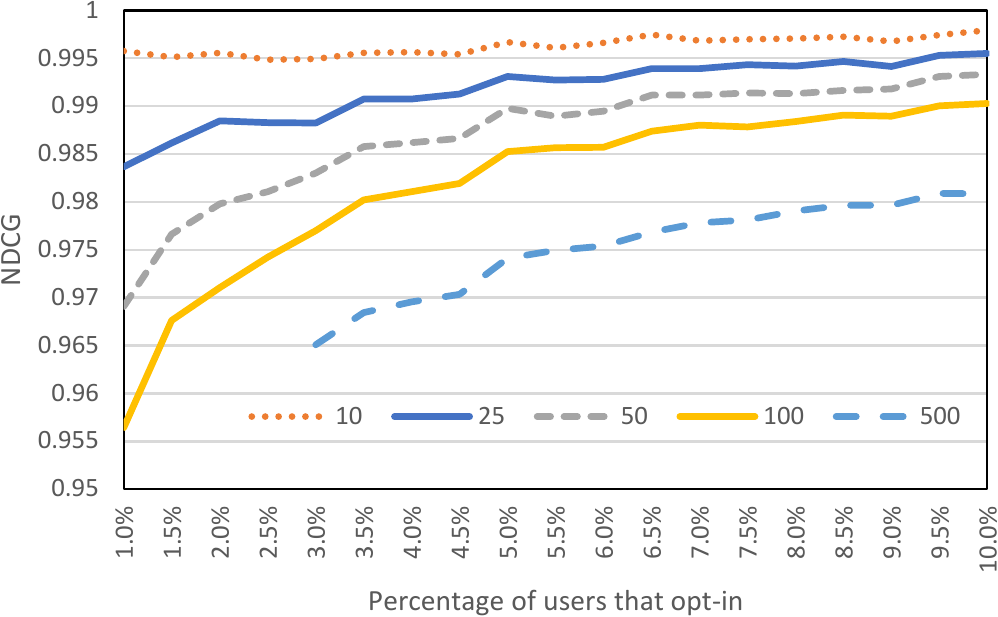}
	\fi
	\caption{Yandex}
\end{subfigure}

\caption{NDCG as a function of the opt-in percentage.}
\label{fig:ndcg-opt-in-percentage}
\end{figure}

\point{Evaluation of local search computation}
Figure~\ref{fig:ndcg-opt-in-percentage} shows the NDCG measurements as a function of the opt-in percentage ranging between~\empirical{1\%} and~\empirical{10\%}.
The results are quite encouraging; for the smaller AOL dataset, we need to have an opt-in level of~\empirical{$\approx$5\%} to achieve an NDCG level in excess of~\empirical{95\%}, which we regard as acceptable. However, for the larger Yandex dataset, we hit that NDCG level even sooner: for an opt-in group composing~\empirical{1\%} of the users, the NDCG level is above~\empirical{95\%} for all but the largest head list size.

\begin{figure}[tb]
\begin{subfigure}{0.49\columnwidth}
\centering
\ifpdf
    \includegraphics[width=.99\columnwidth]{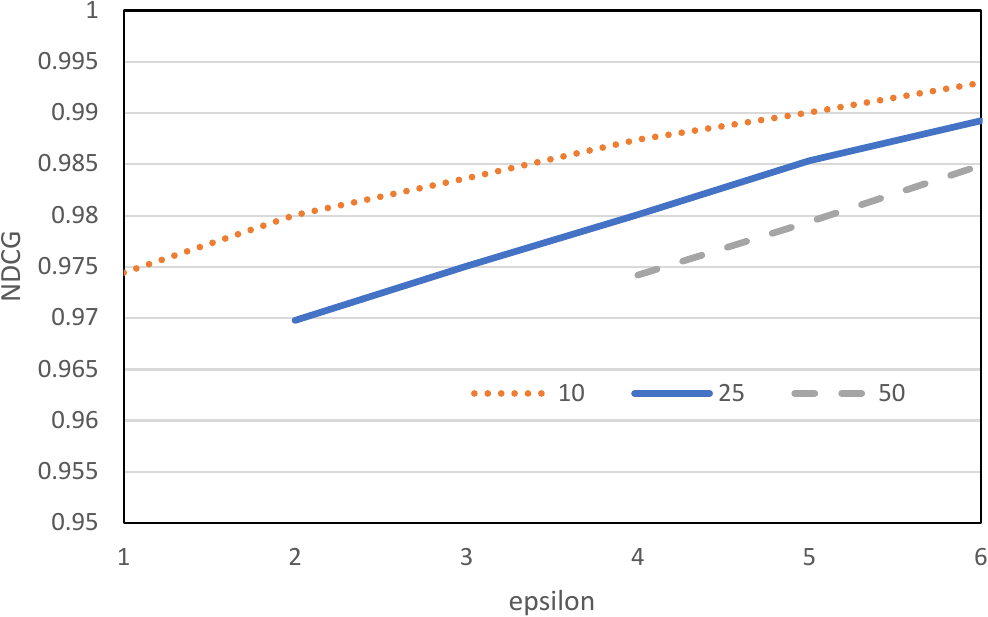}
\fi
\caption{AOL}
\end{subfigure}
\begin{subfigure}{0.49\columnwidth}
\centering
\ifpdf
    \includegraphics[width=.99\columnwidth]{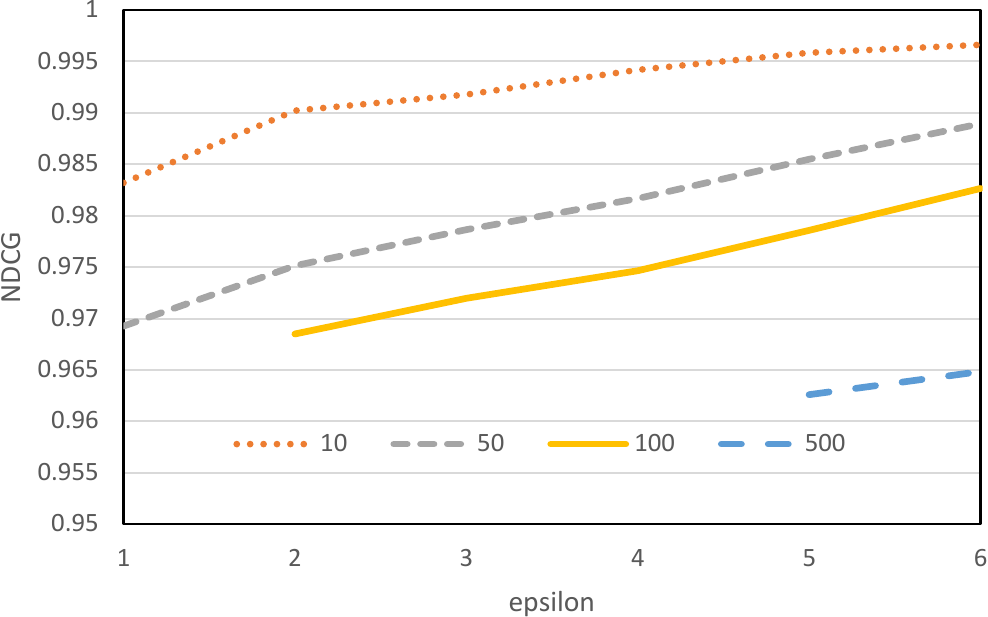}
\fi
\caption{Yandex}
\label{fig:aol-ndcg}
\end{subfigure}
\caption{NDCG statistics for AOL and Yandex datasets for various head list sizes and a range of $\epsilon$ values.}
\label{fig:ndcg-datasets}
\end{figure}

Figure~\ref{fig:ndcg-datasets} shows how the NDCG values vary across the two datasets, AOL and Yandex, for a range of head list sizes and $\epsilon$ values.
We see a clear trend toward higher NDCG values for Yandex, which is not surprising given the sheer volume of data.
For the Yandex dataset, we can keep $\epsilon$ as low as~1 and still achieve NDCG values of~\empirical{95\%} and above for all but the two largest head list sizes. For those, we must increase $\epsilon$ in order to generate larger head lists from the opt-in users.\\

\point{Each group's effect on the blended result}
\begin{figure}[tb]
\begin{subfigure}{0.49\columnwidth}
\centering
\ifpdf
    \includegraphics[width=.99\columnwidth]{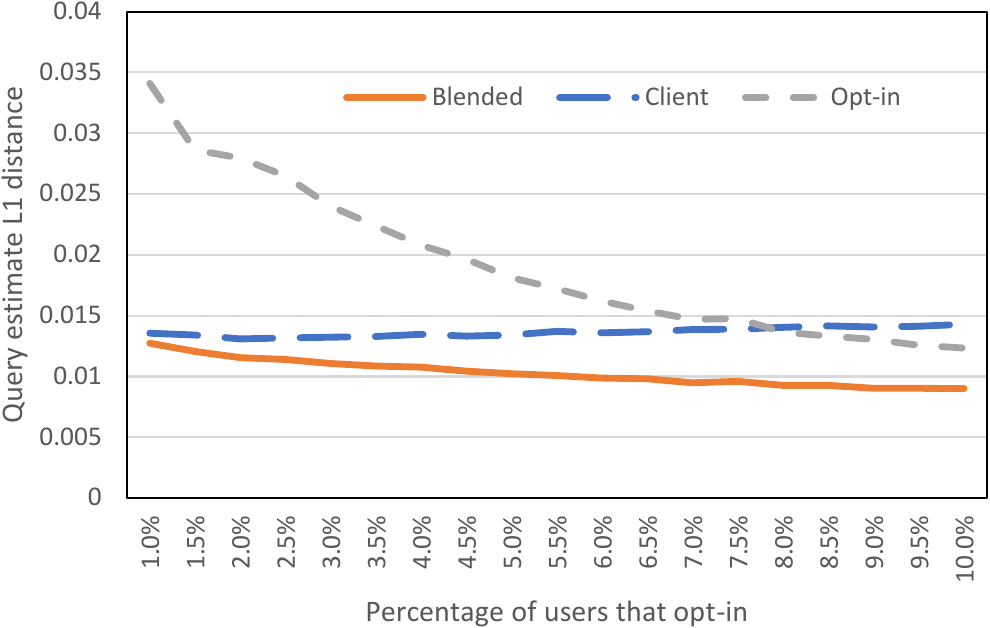}
\fi
\caption{L1 results over opt-in range}
\end{subfigure}
\begin{subfigure}{0.49\columnwidth}
\centering
\ifpdf
    \includegraphics[width=.99\columnwidth]{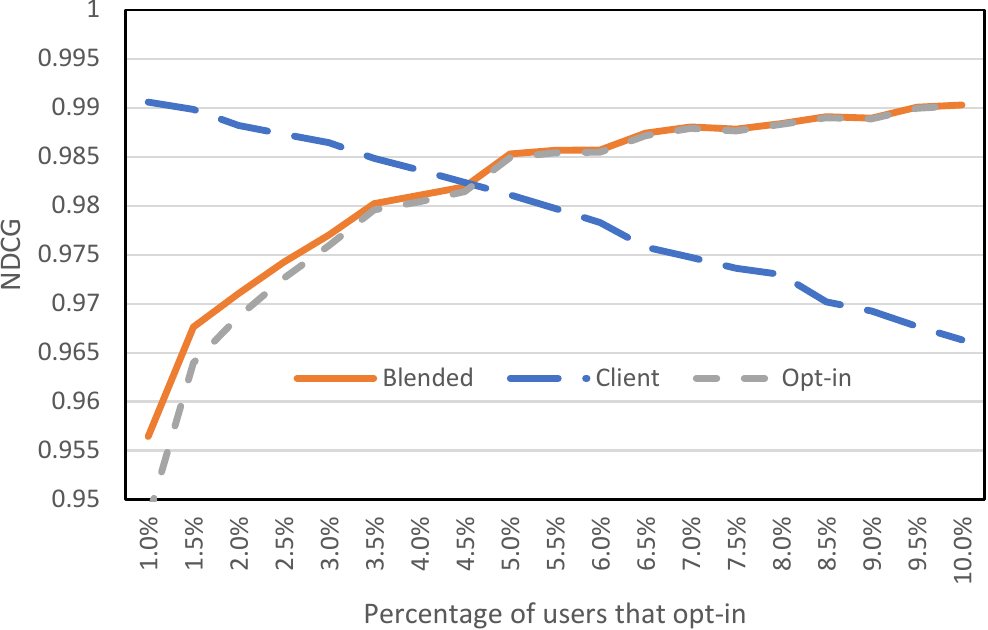}
\fi
\caption{NDCG results over opt-in range}
\end{subfigure}

\vspace{0.25cm}

\begin{subfigure}{0.49\columnwidth}
\centering
\ifpdf
    \includegraphics[width=.99\columnwidth]{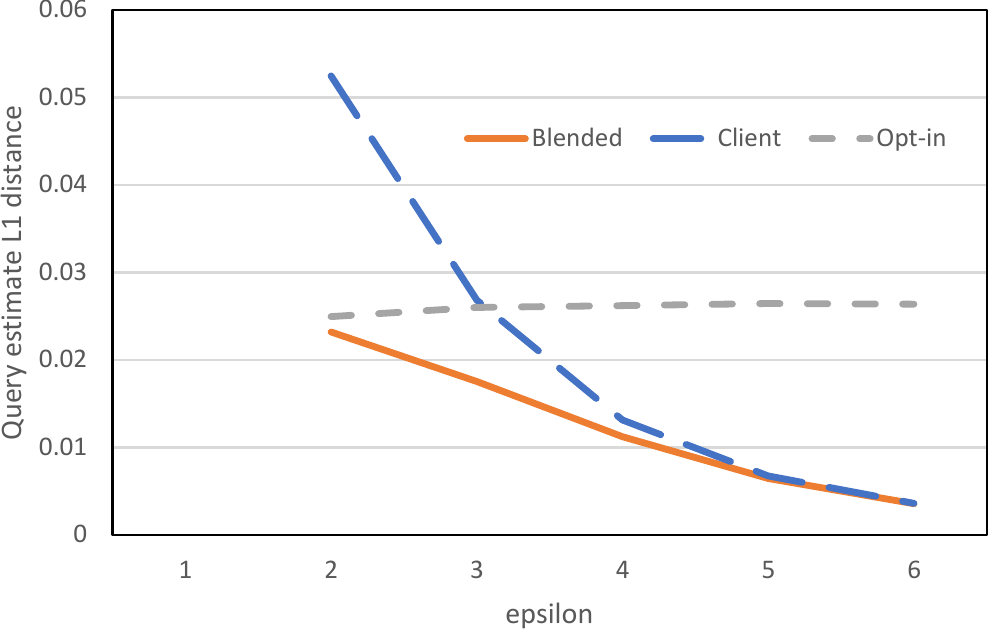}
\fi
\caption{L1 results over $\epsilon$ range}
\end{subfigure}
\begin{subfigure}{0.49\columnwidth}
\centering
\ifpdf
    \includegraphics[width=.99\columnwidth]{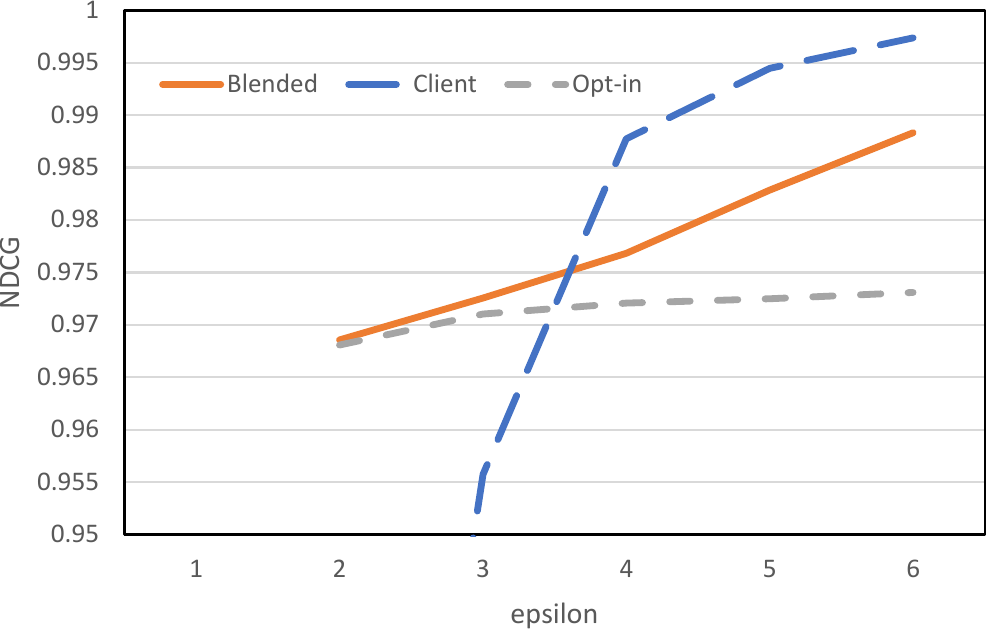}
\fi
\caption{NDCG results over $\epsilon$ range}
\end{subfigure}
\caption{L1 and NDCG statistics broken out between the different groups' results on the Yandex dataset with head list size 100 across a range of opt-in percentages ((a) and (b), with $\epsilon=4$) and a range of $\epsilon$ values ((c) and (d), with 3\% opt-in).}
\label{fig:breakout-stats}
\end{figure}
While these blended results demonstrate the algorithms's high-utility capability, one central question remains: to what extent are each group's estimates contributing to the final blended result?
Specifically, does the small number of samples with low noise from the opt-in group dominate the large number of samples with high noise from the client group, or vice-versa?

For a head list size 100 on the Yandex dataset, Figure~\ref{fig:breakout-stats} examines this question for a range of opt-in percentages and $\epsilon$ values.
These graphs show a complex relationship between the two groups' utility with regards to the final blended result.
In all cases, the blended result is better than the worse of either the opt-in or client results.
With regards to L1 distance, the blended result is better than \emph{both} groups' individual results when varying either the opt-in user percentage or the $\epsilon$ value.

When increasing the opt-in user percentage, the two group's results behave as expected: the opt-in group's results improve as it gains more users, and the client group's results gradually deteriorate as it loses users.
Interestingly, Figures~\ref{fig:breakout-stats}a and b show that the L1 distance of the client group's query results deteriorate quite slowly as their group size decreases, whereas their NDCG results deteriorate more quickly.
To understand this behavior, observe that there are significantly fewer queries (what the query estimate L1 distance is measuring) than there are query-URL pairs (what the NDCG is measuring).
Also note that the utility of the randomized response component of the local algorithm degrades as the set of items under consideration increases.
These two facts in combination explain the difference in the deterioration rates of the client group's utility between Figure~\ref{fig:breakout-stats}a and b.

For the blended result, the NDCG values mainly track the opt-in group's NDCG values even in the case where the client result is clearly better (from \empirical{1\%} up to \empirical{3\%}); this would support the idea that the opt-in results may be dominating the client results when it comes to the blending process.
However, this trend doesn't appear to hold when increasing the $\epsilon$, as the blended results rapidly improve with the client results, while the opt-in results remain relatively flat.
Interestingly, as $\epsilon$ is increased, the opt-in group's L1 results remain relatively constant and its NDCG results only slightly improve.
This is caused by the large amount of noise that is inherent in the opt-in group due to its relatively small size; i.e., a \empirical{3\%} sized opt-in group induces a certain level of sampling error such that the noise introduced for privacy is negligible by comparison.

The takeaway is that there is no single clear-cut group that dominates in its contribution to the final blended result; in fact, both groups appear to contribute across the ranges of parameters considered.\\

\point{When the opt-in group is tiny}
In the real-world, it may be the case that a \empirical{5\%} or even a \empirical{3\%} sized opt-in group is still too large to be considered feasible.
As mentioned in the evaluation of trend computation, the utility results are generally good \emph{conditioned on} the desired head list size being achieved.
\begin{figure}[tb]
\begin{subfigure}{0.49\columnwidth}
\centering
\ifpdf
    \includegraphics[width=.99\columnwidth]{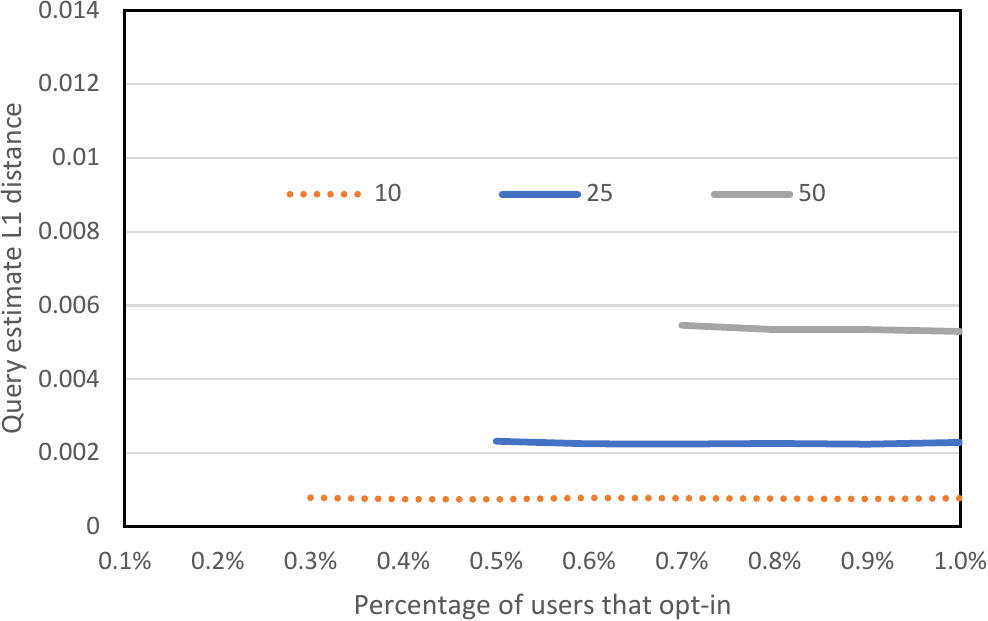}
\fi
\caption{L1 results}
\end{subfigure}
\begin{subfigure}{0.49\columnwidth}
\centering
\ifpdf
    \includegraphics[width=.99\columnwidth]{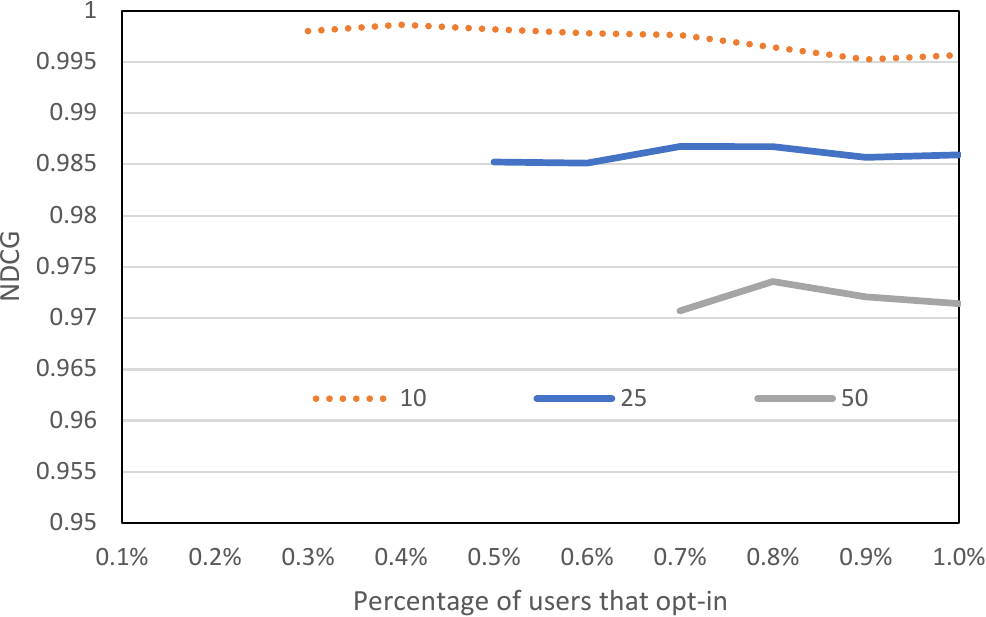}
\fi
\caption{NDCG results}
\end{subfigure}
\caption{L1 and NDCG statistics for the Yandex dataset for various head list sizes across a range of tiny opt-in percentages.}
\label{fig:tiny-optin-stats}
\end{figure}
When the opt-in group becomes too small, it becomes a significantly greater challenge to achieve these large head list sizes.
For the head list sizes that we can achieve at smaller opt-in percentages, what are the utility results we can expect?
Figure~\ref{fig:tiny-optin-stats} shows the performance on the Yandex dataset targeting smaller head list sizes across opt-in group sizes ranging from \empirical{0.1\%} up to \empirical{1\%}.
These results confirm our previous conclusion that once a head list size can be attained, getting high utility probability estimates for the records is a significantly easier challenge.

\begin{figure}[tb]
\begin{subfigure}{0.49\columnwidth}
\centering
\ifpdf
    \includegraphics[width=.99\columnwidth]{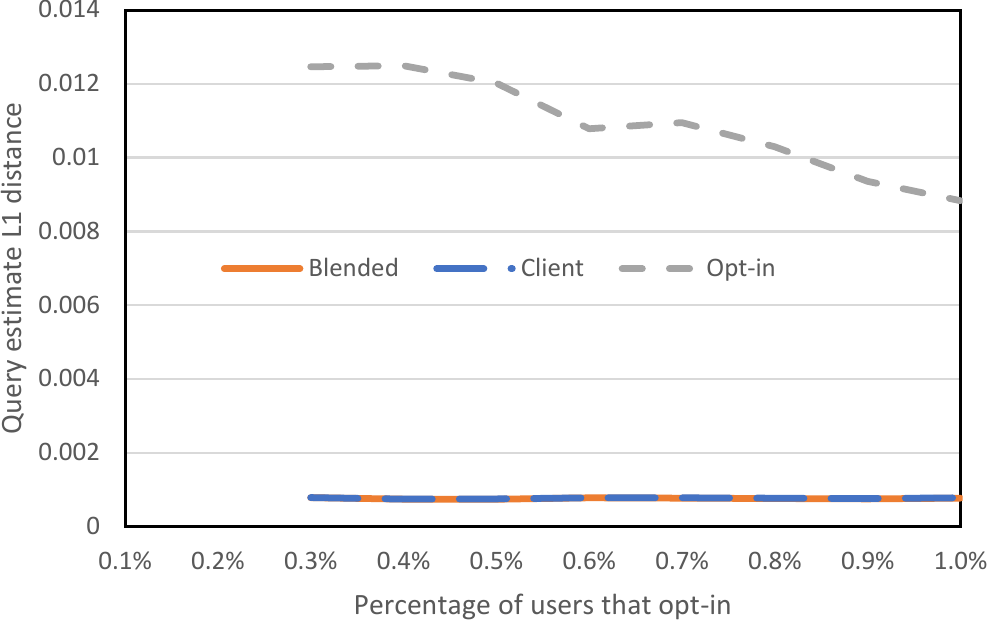}
\fi
\caption{L1 results}
\end{subfigure}
\begin{subfigure}{0.49\columnwidth}
\centering
\ifpdf
    \includegraphics[width=.99\columnwidth]{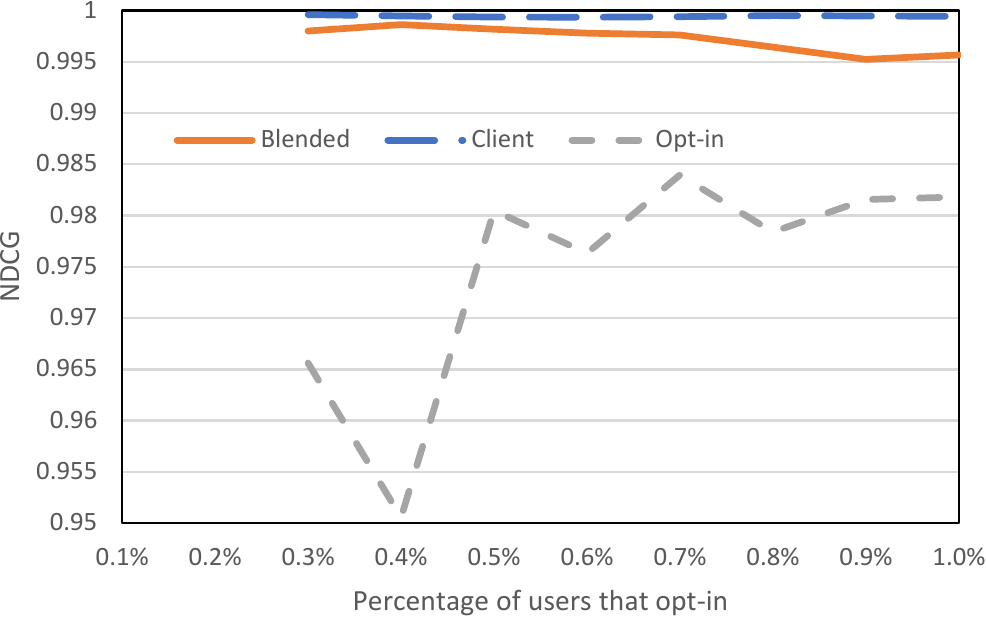}
\fi
\caption{NDCG results}
\end{subfigure}
\caption{L1 and NDCG statistics broken out per group for the Yandex dataset for head list size 10 and a range of tiny opt-in percentages.}
\label{fig:tiny-optin-breakout-stats}
\end{figure}

At these tiny opt-in percentages, with \empirical{95\%} of the opt-in group being assigned to head list creation, only \empirical{0.005\%} to \empirical{0.05\%} of the users are estimating the probabilities under the trusted curator model.
We must ask: in this setting, to what extent \emph{are} the users contributing to the high-utility blended results?
Figure~\ref{fig:tiny-optin-breakout-stats} shows the L1 and NDCG values for the opt-in group, client group, and final blended results across these tiny opt-in sizes for a head list of size 10 on the Yandex dataset.
As suspected, the estimates from the opt-in group have much lower utility relative to the client group.
The blending procedure is able to automatically take advantage of the high variance results of the opt-in group (stemming from the tiny number of samples used by this group in estimating the probabilities) and weigh the blending much more heavily towards the client group's estimates.\\

\section{Related Work}
\label{sec:related}
\point{Algorithms for the trusted curator model}
Researchers have developed numerous privacy-preserving algorithms operating in the trusted curator model that result in useful data for a variety of applications. Specifically, the works of \cite{ korolova2009releasing, gotz2012publishing, korolova2012protecting} pioneered the study of search log data release with differential privacy guarantees; the works of \cite{privbasis} and \cite{bhaskar2010discovering} proposed approaches for privacy-preserving frequent item identification, and so on.\\

\point{Algorithms for the local model}
Although some progress has been made in developing privacy-preserving algorithms operating in the local model~\cite{warner1965randomized, duchi2013local, erlingsson2014rappor, bassily2015local}, the utility of the resulting data is limited~\cite{fanti2016building, kairouz2016discrete}. Furthermore, it is known that for fixed desired differential privacy parameters, the elimination of the trusted data collector comes at the cost of diminished utility~\cite{kairouz2014, klnrs}.
Very recently, much attention has been given to the heavy hitters problem in the local differential privacy setting both from a theoretical and an applied perspective \cite{wang2017ldpheavyhitter, bassily2017practical, jia2018calibrate}.
Since \tool came prior to these recent works, we did not compare our results against theirs\footnote{We note that the new algorithms proposed could be directly applied as the local group's sub-algorithm to improve the utility of \tool.}. \\

\point{Our contribution} Our work significantly improves upon the known results by developing application-specific local model algorithms that work in combination with trusted curator model algorithms.
Specifically, our insight of providing all users with differential privacy guarantees, but achieving it differently depending on whether or not they trust the data curator, enables an efficient privacy-preserving head list construction.
The subsequent usage of this head list in the algorithm operating in the local model helps overcome one of the main challenges to utility of privacy-preserving local algorithms~\cite{fanti2016building}.
As discussed in Section~\ref{sec:ccs}, we significantly outperform previous work of~\cite{qin2016heavy} on metrics of utility in the search context.\\

\section{Conclusions}
\label{sec:conc}

We proposed a hybrid differential privacy model, which permits a mixture of trust models.
In this work, we considered a mix of two primary models studied by the differential privacy community, which differ only in their trust towards the curator: the local model and the trusted curator model.
Using local search as a motivating application, we developed and tested an algorithm which demonstrates that operating in the hybrid model enables significant improvements in terms of utility compared to previously known approaches. Thus, we showed that developing algorithms for hybrid models holds promise for decreasing the gap between theory and practicality of differential privacy.\\

\point{Future work}
The primary direction for future work is to better understand the power of the hybrid model. Specifically, what application areas and algorithms can most effectively utilize data submitted in a mixture of trust models, what utility improvements can such algorithms bring, and how do they depend on the underlying data or user group sizes. Recent work by~\cite{dubey2018power} has begun to study this question for the problem of mean estimation. Works by~\cite{papernot2016semi} and~\cite{feldman2018privacy} show that there is much to be gained by combining trusted curator data with public data, giving another example of a hybrid model that holds promise. A related sub-question is understanding the role that interaction between users contributing data in the local model and in the trusted curator model can play in improving utility.

Another important direction for future work is to address the assumption in current work that user data comes from the same distribution regardless of their trust model, which may not hold in practice. As a start, one can differentially privately evaluate whether the distributions are different using a small sample of records from both groups using the techniques of~\cite{acharya2018differentially, aliakbarpourdr17}. When and how should the differences between groups be taken into account is an open question.

Finally, optimizing the sub-algorithms used by \tool (see Section~\ref{sec:discussion-technical}) and providing \textit{a priori} estimates of its utility under specific assumptions is a promising direction for advancing the deployment of differentially private algorithms in the hybrid model.

\bibliographystyle{alpha}
\bibliography{biblio}

\section{Appendix}
\label{sec:appendix}

\begin{reptheorem}{theorem:localalg-priv}
	\textproc{LocalAlg} is $(\epsilon, \delta)$-differentially private.
\end{reptheorem}

\begin{proof}
We show this by proving that each iteration of the \code{for} loop in line~\ref{local-loop} of \textproc{LocalAlg} is $(\epsilon^\prime, \delta^\prime)$-differentially private, where $\epsilon^\prime = \epsilon/m_C$ and $\delta^\prime = \delta/ m_C$.
Since there are at most $m_C$ iterations of this loop for each client, composition of differentially private algorithms~\cite{dwork2010boosting} guarantees that \textproc{LocalAlg} ensures $(\epsilon, \delta)$-differential privacy for each client.

Denote each iteration of the \code{for} loop in line~\ref{local-loop} of \textproc{LocalAlg} by $L$; it takes as input a record $\langle q,u\rangle\in D$, and returns a record, which we denote $L(\langle q,u\rangle)$.
If $q$ is not in $HL$ or $u$ is not in $HL[q],$ then they immediately get transformed into a default value $(\star)$ that is in the head list.
Since $L$ outputs only values that exist in the head list, to confirm differential privacy we need to prove that for any arbitrary neighboring datasets $\langle q,u\rangle$ and $\langle q^\prime,u^\prime\rangle$,  $\Pr\bigl[L(\langle q,u\rangle) \in Y\bigr] \le e^{\epsilon^\prime} \Pr\bigl[L(\langle q^\prime,u^\prime\rangle) \in Y\bigr] + \delta^\prime$ holds for all sets of head list records $Y$.

Whenever $k=1$ or $k_q=1$, the only query (or URL for a specific query) is $\star$, which will be output with probability 1.
Thus, differential privacy trivially holds, since the reported values then do not rely on the client's data.
Thus, we'll assume $k \ge 2$ and $k_q \ge 2$.
Note that there is a single decision point where it is determined whether $q$ will be reported truthfully or not.
Thus, we can split the privacy analysis into two parts: 1) Usage of the $f_C$ fraction of the privacy budget to report a query, and 2) Usage of the remainder of the privacy budget to report a URL (given the reported query).
This decomposes a simultaneous two-item $(\epsilon^\prime, \delta^\prime)$ reporting problem into two single-item reporting problems with $(\epsilon^\prime_Q, \delta^\prime_Q)$ and $(\epsilon'_U, \delta'_U)$ respectively, where $\epsilon^\prime_Q = f\epsilon^\prime,\ \delta^\prime_Q = f\delta^\prime,\  \epsilon^\prime_U = (1-f_C)\epsilon^\prime,\ $ and $\delta^\prime_U = (1-f_C)\delta'$.

\point{1. Privacy of query reporting}
Consider the query-reporting case first.
Overloading our use of $L$, let $L(q)$ be the portion of $L$ that makes use of $q$.
We first ensure that
\footnotesize
\begin{align}
\Pr[L(q)=q_{HL}] \le \exp(\epsilon^\prime_Q)\Pr[L(q^\prime)=q_{HL}] + \frac{\delta^\prime_Q}{2} \label{eq:dp-singleton}
\end{align}
\normalsize
holds for all $q,q^\prime,$ and $q_{HL} \in HL$.
This trivially holds when $q_{HL}=q=q^\prime$ or $q_{HL} \not \in \{q,q^\prime\}$.
The remaining scenarios to consider are: 1) $q \neq q_{HL}, q^\prime = q_{HL}$ and 2) $q = q_{HL}, q^\prime \neq q_{HL}$.
By the design of the algorithm,
$\Pr[L(q_{HL})=q_{HL}] = t$
and
$\Pr[L(\bar{q}_{HL})=q_{HL}]=(1-t)(\frac{1}{k-1}),$
where $\bar{q}_{HL}$ represents any query not equal to $q_{HL}$.
With
$t = \frac{\exp(\epsilon^\prime_Q) + (\delta^\prime_Q /2)(k-1)}{\exp(\epsilon^\prime_Q)+k-1},$
it is simple to verify that inequality~(\ref{eq:dp-singleton}) holds.

Consider an arbitrary set of head list queries $Y$.
\footnotesize
\begin{align}
\Pr[L(q) \in Y] &= \smashoperator[r]{\sum_{q_{HL} \in Y}} \Pr[L(q) = q_{HL}] \notag \\
&= \smashoperator[r]{\sum_{q_{HL} \in Y \setminus \{q,q^\prime\}}} \Pr[L(q) = q_{HL}] + \smashoperator{\sum_{q_{HL} \in Y \cap \{q,q^\prime\}}} Pr[L(q) = q_{HL}] \notag \\
&= \smashoperator[r]{\sum_{q_{HL} \in Y \setminus \{q,q^\prime\}}} \Pr[L(q^\prime) = q_{HL}] + \smashoperator{\sum_{q_{HL} \in Y \cap {q,q^\prime}}} \Pr[L(q) = q_{HL}] \label{eq:dp-notneighbor} \\
&\le \smashoperator[r]{\sum_{q_{HL} \in Y \setminus \{q,q^\prime\}}} \Pr[L(q^\prime) = q_{HL}] + \smashoperator{\sum_{q_{HL} \in Y \cap \{q,q^\prime\}}} \bigl(e^{\epsilon^\prime_Q}\Pr[L(q') = q_{HL}] + \frac{\delta^\prime_Q}{2}\bigr) \label{eq:dp-neighbor} \\
&\le e^{\epsilon_Q^\prime}\smashoperator[r]{\sum_{q_{HL} \in Y}} \Pr[L(q^\prime) = q_{HL}] + 2\cdot\frac{\delta^\prime_Q}{2} \notag \\
&= e^{\epsilon^\prime_Q}\Pr[L(q^\prime) \in Y] + \delta^\prime_Q, \notag
\end{align}
\normalsize
Equality~(\ref{eq:dp-notneighbor}) stems from the fact that the probability of reporting a false query is independent of the user's true query. The inequality~(\ref{eq:dp-neighbor}) is a direct application of inequality~(\ref{eq:dp-singleton}).
Thus, $L$ is $(\epsilon_Q^\prime, \delta_Q^\prime)$-differentially private for query-reporting.

\point{2. Privacy of URL reporting}
With $t_q$ defined as
$t_q = \frac{\exp(\epsilon^\prime_U) + 0.5\delta^\prime_U (k_q-1)}{\exp(\epsilon^\prime_U) + k_q-1},$
an analogous argument shows that the $(\epsilon_U^\prime, \delta_U^\prime)$-differential privacy constraints hold if the original $q$ is kept.
On the other hand, if it is replaced with a random query, then they trivially hold as the algorithm reports a random element in the URL list of the reported query, without taking into consideration the client's true URL $u$.

By composition~\cite{dwork2010boosting}, each of the at most $m_C$ iterations of $L$ is $(\epsilon^\prime_Q + \epsilon^\prime_U, \delta^\prime_Q + \delta^\prime_U) = (\epsilon^\prime, \delta^\prime)$-differentially private.
\end{proof}

\vspace{1cm}

\begin{repobservation}{observation:client-denoising}
$\hat{p}_C$ gives the unbiased estimate of record and query probabilities under \textproc{EstimateClientProbabilities}.
\end{repobservation}

\begin{proof}
Reporting records is a two-stage process (first, decide which query to report, then report a record); similarly, denoising is also done in two stages.

\point{Denoising of query probability estimates}
Let $r_{C,q}$ denote the probability that the algorithm has received query $q$ as a report, and let $p_{q}$ be the true probability of a user having query $q$. We want to learn $p_q$ based on $r_{C,q}$.
By the design of our algorithm,
\footnotesize
\begin{align*}
r_{C,q} &= t\cdot p_{q} + \sum_{q' \neq q}p_{q'}(1-t)\frac{1}{k-1}\\
&= t\cdot p_{q} + \frac{1-t}{k-1}\sum_{q' \neq q}p_{q'}\\
&= t\cdot p_{q} + \frac{1-t}{k-1}(1-p_{q}).
\end{align*}
\normalsize

Solving for $p_q$ in terms of $r_{C,q}$ yields
$p_{q} = \frac{r_{C,q} - \frac{1-t}{k-1}}{t - \frac{1-t}{k-1}}.$
Using the obtained data for the query $\hat{r}_{C,q}$, we estimate $p_{C,q}$ as
$\hat{p}_{C,q} = \frac{\hat{r}_{C,q} - \frac{1-t}{k-1}}{t - \frac{1-t}{k-1}}.$

\point{Denoising of record probability estimates}
Analogously, denote by $r_{C,\langle q,u\rangle}$ the probability that the algorithm has received a record $\langle q,u\rangle$ as a report, and recall $p_{\langle q,u\rangle}$ is the record's true probability in the dataset. Then 
$r_{C,\langle q,u\rangle} = t\cdot t_q\cdot p_{\langle q,u\rangle} + \bigl(t \frac{1 - t_q}{k_q - 1}\bigr)(p_q - p_{\langle q,u\rangle}) +
    \bigl(\frac{1 - t}{k - 1}\cdot \frac{1}{k_q}\bigr)(1 - p_q)$,
\normalsize
recalling from the algorithm that $k_q$ is the number of URLs associated with query $q$ and $t_q$ is the probability of truthfully reporting $u$ given that query $q$ was reported.
Solving for $p_{\langle q,u\rangle}$ yields
$p_{\langle q,u\rangle} = \frac{r_{C,\langle q,u\rangle} - \bigl(t\frac{1-t_q}{k_q-1}p_q + \frac{(1-t)(1-p_q)}{(k-1)k_q}\bigr)}{t(t_q - \frac{1-t_q}{k_q-1})}.$

Using the obtained data for the empirical report estimate $\hat{r}_{C,\langle q,u\rangle}$ together with the query estimate $\hat{p}_{C,q}$, we estimate $p_{\langle q,u\rangle}$ as
$\hat{p}_{C,\langle q,u\rangle} = \frac{\hat{r}_{C,\langle q,u\rangle} - \bigl(t\frac{1-t_q}{k_q-1}\hat{p}_{C,q} + \frac{(1-t)(1-\hat{p}_{C,q})}{(k-1)k_q}\bigr)}{t(t_q - \frac{1-t_q}{k_q-1})}.$
\end{proof}

\vspace{1cm}
\begin{reptheorem}{theorem:opt-in-variance}
If $m_O = 1$ then the unbiased variance estimate for the opt-in group's record probabilities can be computed as:
$\hat{\sigma}^2_{O,\langle q,u\rangle} = \frac{|D_T|}{|D_T|-1}\left(\frac{\hat{p}_{O,\langle q,u\rangle}(1-\hat{p}_{O,\langle q,u\rangle})}{|D_T|} + 2 \left(\frac{b_T}{|D_T|}\right)^2\right).$
\end{reptheorem}

\begin{proof}
Given the head list, the distribution of \textproc{EstimateOptinProbabilities}' estimate for a record $\langle q,u\rangle$ is given by $r_{O,\langle q,u\rangle} = p_{\langle q,u\rangle} + \frac{Y}{|D_T|}$, where $Y \sim \textrm{Laplace}(b_T)$ with $b_T$ being the scale parameter and recalling that $|D_T|$ is the total number of records from the opt-in users used to estimate probabilities.
The empirical estimator for $r_{O,\langle q,u\rangle}$ is $\hat{r}_{O,\langle q,u\rangle} = \frac{1}{|D_T|}\sum_{j=1}^{|D_T|}X_j + Y$, where $X_j \sim \textrm{Bernoulli}(p_{\langle q,u\rangle})$ is the random variable indicating whether report $j$ was record $\langle q,u\rangle$.

The expectation of this estimator is given by
$\E[\hat{r}_{O,\langle q,u\rangle}] = p_{\langle q,u\rangle}$. Thus, $\hat{r}_{O,\langle q,u\rangle}$ is an unbiased estimator for $p_{\langle q,u\rangle}$.
We denote $
\hat{p}_{O,\langle q,u\rangle} = \hat{r}_{O,\langle q,u\rangle}$ to explicitly reference it as the estimator of $p_{\langle q,u\rangle}$.
The variance for this estimator is
\footnotesize
\begin{align}
\sigma^2_{O,\langle q,u\rangle} &= \Var[\hat{p}_{O,\langle q,u\rangle}] \\
&= \Var\Bigl[\frac{1}{|D_T|}\bigl(\sum_{j=1}^{|D_T|}X_j + Y\bigr)\Bigr] \notag \\
&= \frac{1}{|D_T|^2}\Bigl(\Var\bigl[\sum_{j=1}^{|D_T|}X_j\bigr] + \Var\left[Y\right]\Bigr) \label{eq:var-indep1} \\
&= \frac{1}{|D_T|^2}\Bigl(\sum_{j=1}^{|D_T|}\Var\left[X_j\right] + \Var\left[Y\right]\Bigr) \label{eq:var-indep2} \\
&= \frac{1}{|D_T|^2}\bigl(|D_T|\cdot p_{\langle q,u\rangle}(1-p_{\langle q,u\rangle})\bigr) + 2\Bigl(\frac{b_T}{|D_T|}\Bigr)^2 \notag \\
&= \frac{p_{\langle q,u\rangle}(1-p_{\langle q,u\rangle})}{|D_T|} + 2\Bigl(\frac{b_T}{|D_T|}\Bigr)^2. \notag
\end{align}
\normalsize
Equality~\ref{eq:var-indep1} comes from the independence between $Y$ and all $X_j$.
Equality~\ref{eq:var-indep2} relies on an assumption of independence between $X_j, X_k$ for all $j \neq k$ (i.e., the iid assumption discussed prior to the theorem statements).

To compute this variance, we need to use the data in place of the unknown $p_{\langle q,u\rangle}$.
Using $\hat{p}_{O,\langle q,u\rangle}$ directly in place of $p_{\langle q,u\rangle}$ requires a $\frac{|D_T|}{|D_T|-1}$ factor correction (known as ``Bessel's correction\footnote{\url{https://en.wikipedia.org/wiki/Bessel's_correction}}'') to generate an unbiased estimate. Thus, the variance of each opt-in record probability estimate is:
$\hat{\sigma}^2_{O,\langle q,u\rangle} = \frac{|D_T|}{|D_T|-1}\left(\frac{\hat{p}_{O,\langle q,u\rangle}(1-\hat{p}_{O,\langle q,u\rangle})}{|D_T|} + 2\left(\frac{b_T}{|D_T|}\right)^2\right).$
\end{proof}

\vspace{1cm}
\begin{reptheorem}{theorem:client-variance}
If $m_C = 1$ then the unbiased variance estimate for the client group's record probabilities can be computed as:
\footnotesize
\begin{align*}
\hat{\sigma}^2_{C,\langle q,u\rangle}& = \frac{|D_C|}{t^2\bigl(t_q - \frac{1-t_q}{k_q-1}\bigr)^2 (|D_C|-1)} \cdot\\
		&\Bigl( \frac{\hat{r}_{C,\langle q,u\rangle}(1-\hat{r}_{C,\langle q,u\rangle})}{|D_C|}
			+ \bigl(\frac{1-t}{(k-1)k_q} - t\frac{1-t_q}{k_q-1} \bigr)^2 \hat{\sigma}^2_{C,q} + 2 
			\bigl(\frac{1-t}{(k-1)k_q} -~t\frac{1-t_q}{k_q-1} \bigr)\frac{\hat{r}_{C,\langle q,u\rangle}(1-\hat{r}_{C,q})}{|D_C| (t-\frac{1-t}{k-1})} \Bigr).
\end{align*}
\normalsize
\end{reptheorem}

\begin{proof}
We'll first derive the variance estimate for the client group's query probabilities, then move on to the variance estimate for their record probabilities.

From the proof of Observation~\ref{observation:client-denoising}, the distribution of the reported query $q$ from the client algorithm is given by $r_{C,q} = t\cdot p_q + \frac{1-t}{k-1}(1-p_q)$, and so the true probability of query $q$ is distributed as $p_q = \frac{r_{C,q} - \frac{1-t}{k-1}}{t - \frac{1-t}{k-1}}$.
The empirical estimator for $p_q$ is $\hat{p}_{C,q} = \frac{\hat{r}_{C,q} - \frac{1-t}{k-1}}{t - \frac{1-t}{k-1}}$, where $\hat{r}_{C,q}$ is the empirical estimator of $r_{C,q}$ defined explicitly as $\hat{r}_{C,q} = \frac{1}{|D_C|}\sum_{j=1}^{|D_C|}X_j$, where $X_j \sim \textrm{Bernoulli}(r_{C,q})$ is the random variable indicating whether report $j$ was query $q$ and recalling that $|D_C|$ is the total number of records from the client users.

The variance of $\hat{r}_{C,q}$ is
\footnotesize
\begin{align}
\Var[\hat{r}_{C,q}] &= \Var\Bigl[\frac{1}{|D_C|}\sum_{j=1}^{|D_C|}X_j\Bigr] \notag \\
&= \Bigl(\frac{1}{|D_C|}\Bigr)^2\sum_{j=1}^{|D_C|}\Var\left[X_j\right] \label{eq:var-indep3} \\
&= \bigl(\frac{1}{|D_C|}\bigr)^2 \bigl(|D_C|\cdot r_{C,q}(1-r_{C,q})\bigr) \\
&= \frac{r_{C,q}(1-r_{C,q})}{|D_C|}, \notag
\end{align}
\normalsize
where equality~\ref{eq:var-indep3} relies on an assumption of independence between $X_j, X_k$ for all $j \neq k$ (i.e., the iid assumption discussed prior to the theorem statements).

Then, the variance of $\hat{p}_{C,q}$ is
\footnotesize
$$\sigma^2_{C,q} = \Var[\hat{p}_{C,q}] = \Var\Bigl[\frac{\hat{r}_{C,q} - \frac{1-t}{k-1}}{t - \frac{1-t}{k-1}} \Bigr] = \frac{r_{C,q}(1-r_{C,q})}{|D_C| \bigl(t - \frac{1-t}{k-1}\bigr)^2}.$$
\normalsize

To compute this variance, we need to use the data in place of the unknown $r_{C,q}$.
Using $\hat{r}_{C,q}$ directly in place of $r_{C,q}$ requires including Bessel's $\frac{|D_C|}{|D_C|-1}$ factor correction to yield an unbiased estimate. Thus, the variance of the query probability estimates by the client algorithm is:
$\hat{\sigma}^2_{C,q} = \left(\frac{1}{t - \frac{1-t}{k-1}}\right)^2 \frac{\hat{r}_{C,q}(1-\hat{r}_{C,q})}{|D_C|-1}.$\\

Now, we'll derive the variance estimate for the record probabilities.
For a given query $q$ and corresponding URL $u$ in head list, denote $X_i^q$ as the indicator random variable that is 1 if user $i$ reported query $q$ and 0 otherwise, and similarly denote $X_i^{\langle q,u\rangle}$ as the indicator random variable that is 1 if user $i$ reported query $q$ and URL $u$ and 0 otherwise.
Note that $X_i^q \sim$ Bern$(r_{C,q})$ and $X_i^{\langle q,u\rangle} \sim$ Bern$(r_{C,\langle q,u\rangle})$.
The covariance between these two random variables is given by
\footnotesize
$$\Cov[X_i^q, X_i^{\langle q,u\rangle}] = \E[X_i^q X_i^{\langle q,u\rangle}] - \E[X_i^q] \E[X_i^{\langle q,u\rangle}] = r_{C,\langle q,u\rangle} - r_{C,
\langle q,u\rangle} r_{C,q} = r_{C,\langle q,u\rangle} (1 - r_{C,q}).$$
\normalsize
Also note that due to the iid assumption, for any other user $j$, we have $\Cov(X_i^q, X_j^{\langle q,u\rangle}) = 0$.
Thus, we have the covariance between our empirical query and record estimates as
\footnotesize
\begin{align*}
\Cov[\hat{r}_q, \hat{r}_{\langle q,u\rangle}] &= \Cov\left[\frac{1}{|D_C|} \sum_{i\in D_C} X_i^q, \frac{1}{|D_C|} \sum_{i\in D_C} X_i^{\langle q,u\rangle}\right] \\
&= \frac{1}{|D_C|^2} \Cov\left[\sum_{i\in D_C} X_i^q, \sum_{i\in D_C} X_i^{\langle q,u\rangle}\right] \\
&= \frac{1}{|D_C|^2} \sum_{i,j \in D_C} \Cov[X_i^q, X_j^{\langle q,u\rangle}] \\
&= \frac{1}{|D_C|^2} \sum_{i \in D_C} \Cov[X_i^q, X_i^{\langle q,u\rangle}] \\
&= \frac{r_{C,\langle q,u\rangle} (1-r_{C,q})}{|D_C|}.
\end{align*}
\normalsize

Utilizing this covariance expression, we can now compute the desired variance estimate as:
\footnotesize
\begin{align*}
&\sigma^2_{C, \langle q,u\rangle} = \Var[\hat{p}_{C, \langle q,u\rangle}] \\
&= \Var\left[ \frac{\hat{r}_{C,\langle q,u\rangle} - \bigl(t\frac{1-t_q}{k_q-1}\hat{p}_{C,q} + \frac{(1-t)(1-\hat{p}_{C,q})}{(k-1)k_q}\bigr)}{t(t_q - \frac{1-t_q}{k_q-1})} \right] \\
&= \frac{1}{t^2(t_q - \frac{1-t_q}{k_q-1})^2} \Var\left[\hat{r}_{C,\langle q,u\rangle} - \bigl(t\frac{1-t_q}{k_q-1}\hat{p}_{C,q} + \frac{(1-t)(1-\hat{p}_{C,q})}{(k-1)k_q}\bigr) \right] \\
&= \frac{1}{t^2(t_q - \frac{1-t_q}{k_q-1})^2} \Var\left[\hat{r}_{C,\langle q,u\rangle} - \hat{p}_{C,q} \bigl(\frac{1-t}{(k-1)k_q} - t\frac{1-t_q}{k_q-1}\bigr) \right] \\
&= \frac{1}{t^2(t_q - \frac{1-t_q}{k_q-1})^2} \cdot \\
&\quad \left(\Var\left[\hat{r}_{C,\langle q,u\rangle}\right] + \bigl(\frac{1-t}{(k-1)k_q} - t\frac{1-t_q}{k_q-1}\bigr)^2 \Var\left[\hat{p}_{C,q}\right] + 2\bigl(\frac{1-t}{(k-1)k_q} - t\frac{1-t_q}{k_q-1}\bigr)\Cov[\hat{p}_{C,q}, \hat{r}_{C,\langle q,u\rangle}]\right) \\
&= \frac{1}{t^2(t_q - \frac{1-t_q}{k_q-1})^2} \cdot \\
&\quad \left(\frac{r_{C,\langle q,u\rangle}(1-r_{C,\langle q,u\rangle})}{|D_C|} + \bigl(\frac{1-t}{(k-1)k_q} - t\frac{1-t_q}{k_q-1}\bigr)^2 \sigma^2_{C,q} + 2\bigl(\frac{1-t}{(k-1)k_q} - t\frac{1-t_q}{k_q-1}\bigr)\frac{1}{t-\frac{1-t}{k-1}}\Cov[\hat{r}_{C,q}, \hat{r}_{C,\langle q,u\rangle}]\right) \\
&= \frac{1}{t^2(t_q - \frac{1-t_q}{k_q-1})^2} \cdot \\
&\quad \left(\frac{r_{C,\langle q,u\rangle}(1-r_{C,\langle q,u\rangle})}{|D_C|} + \bigl(\frac{1-t}{(k-1)k_q} - t\frac{1-t_q}{k_q-1}\bigr)^2 \sigma^2_{C,q} + 2\bigl(\frac{1-t}{(k-1)k_q} - t\frac{1-t_q}{k_q-1}\bigr)\frac{1}{t-\frac{1-t}{k-1}} \frac{r_{C,\langle q,u\rangle} (1-r_{C,q})}{|D_C|}\right).
\end{align*}
\normalsize
Using our already-computed estimates $\hat{r}_{C,q}, \hat{r}_{C,\langle q,u\rangle},$ and $\hat{\sigma}^2_{C,\langle q,u\rangle}$ (in place of $r_{C,q}, r_{C,\langle q,u\rangle},$ and $\sigma^2_{C,\langle q,u\rangle}$ respectively) and applying Bessel's correction, we obtain the stated result.
\end{proof}

\vspace{1cm}
\begin{reptheorem}{theorem:weighting-scheme}
If $\hat{\sigma}^2_{O,\langle q,u\rangle}$ and $\hat{\sigma}^2_{C,\langle q,u\rangle}$ are sample variances of $\hat{p}_{O,\langle q,u\rangle}$ and $\hat{p}_{C, \langle q,u\rangle}$ respectively, and the blended estimate is the convex combination $\hat{p}_{\langle q,u\rangle} = w_{\langle q,u\rangle}\cdot\hat{p}_{O,\langle q,u\rangle} + (1-w_{\langle q,u\rangle})\cdot\hat{p}_{C,\langle q,u\rangle}$, then the sample variance optimal weighting is given by $w_{\langle q,u\rangle} = \frac{\hat{\sigma}^2_{C,\langle q,u\rangle}}{\hat{\sigma}^2_{O,\langle q,u\rangle} + \hat{\sigma}^2_{C,\langle q,u\rangle}}$.
\end{reptheorem}

\begin{proof}
With the record probability and variance estimates for each group fully computed, the blended estimate of $p_{\langle q,u\rangle}$ is given by
$\hat{p}_{\langle q,u\rangle} = w_{\langle q,u\rangle}\cdot\hat{p}_{O,\langle q,u\rangle} + (1-w_{\langle q,u\rangle})\cdot\hat{p}_{C,\langle q,u\rangle}$.
The sample variance of $\hat{p}_{\langle q,u\rangle}$  is given by $\hat{\sigma}^2_{\langle q,u\rangle} = w_{\langle q,u\rangle}^2\cdot\hat{\sigma}^2_{O,\langle q,u\rangle} + (1-w_{\langle q,u\rangle})^2\cdot\hat{\sigma}^2_{C,\langle q,u\rangle}.$
Minimizing $\hat{\sigma}^2_{\langle q,u\rangle}$ with respect to $w_{\langle q,u\rangle}$ yields the stated result.
\end{proof}

\end{document}